\documentclass[preprint,12pt]{elsarticle}
\pdfoutput=1 
\usepackage{graphicx}
\usepackage{amsfonts}
\usepackage{amsmath}
\usepackage{url}
\usepackage{comment}
\usepackage{epstopdf}
\usepackage{hyperref} 
\usepackage{color}
\usepackage{rotating}

\def\ve{\varepsilon}
\def\C{{\mathbf C}}
\def\A{{\mathbf A}}
\def\S{{\mathbf S}}
\def\E{{\mathbf E}}
\def\M{{\mathbf M}}
\def\U{{\mathbf U}}

\def\Nu{{\mathcal V}}

\def\P{{\mathcal P}}

\def\var{\mathrm{var}}

\def\T{\dagger}

\def\s{{\bf s}}

\def\var{{\rm var}}
\def\PNL{{\mathcal P}}  
\def\dPNL{\delta\PNL}  
\def\Sh{{\mathcal S}}

\begin{document}

\begin{frontmatter}
\title{Optimal trend following portfolios}

\author{Sebastien Valeyre}
\address{
Sebastien Valeyre is a senior manager at  Valeyre Research, 114 rue de Antibes, 06400 Cannes, France, Phone +33 3 22 21 65 96. Email: sb.valeyre@gmail.com}
\ead{sb.valeyre@gmail.com}



\date{\today}

\begin{abstract}
	This paper derives an optimal portfolio that is based on trend-following signal.
	Building on an earlier related article,  it provides a unifying theoretical setting to introduce an autocorrelation model with the covariance matrix of trends and risk premia.
	We specify practically relevant models for the covariance matrix of
	trends. The optimal portfolio is decomposed into four basic
components that yield four basic portfolios: Markowitz, risk parity,
agnostic risk parity, and trend following on risk parity.  The
overperformance of the proposed optimal portfolio, applied to
cross-asset trading universe, is confirmed by empirical backtests.  We
provide thus a unifying framework to describe and rationalize earlier
developed portfolios.
\end{abstract}


\end{frontmatter}

\newpage

\section{Introduction}

In systematic trading on exchange markets, detecting minuscule trends
in asset price fluctuations is like looking for a needle in a
haystack.  While the chance of a correct forecast of the next price
move is fairly close to $50\%$, even a minor excess above this value
can lead to significant profits after multiple transactions supervised
by an algoritmic trading system according to prescribed rules.  
A key feature of trend following investment strategies is that they
can be implemented by applying simple rules 
\cite{Clare17}.
For instance, trend following (TF) strategies adjust their market
exposure by {\it assuming} the next move of an asset price to be in a
trend with its past variations
\cite{Covel,Clenow,Fung01,Clare12,Asness13}.  While its
profitability is debatable (as it contradicts the market efficiency
hypothesis) \cite{Potters05,Martin12,Chan96,Jegadeesh01,Moskowitz12},
trend following remains a widely used strategy among professional
asset managers.

Most explanations on trend following success rely on behavioral
theory of asset pricing, which includes boundedly rational investors'
initial under-reaction to new information that allows momentum traders
to take profit of any under-reaction by trend chasing \cite{Hong99}.
It also includes well known herding behavior, which gives rise to a
collective decision-making process by investors' beliefs that are
either optimistic or pessimistic \cite{Barberis03}.  More intuitively,
Bhansali {\it et al.} explain the ability of trend following to
deliver substantial returns because it is a cousin of the
cross-sectional momentum anomaly \cite{Bhansali15}.  Hence, a
trend-following strategy typically takes long positions in securities
with positive past returns and short positions in securities with
negative past returns.  As such, Moskowitz {\it et al.} document that
a portfolio of time-series momentum strategies, or equivalently
trend-following strategies, across all asset classes delivers
substantial abnormal returns performs best during extreme markets
\cite{Moskowitz12}.  In the same vein, Hurst {\it et al.} find that
trend following has been consistently profitable throughout the past
137 years, which makes them conclude that based on their long-term
out-of-sample evidence that price trends in markets is not a product
of statistical randomness or even data mining \cite{Hurst17}.
Even though forecasting individual asset prices is rather hopeless,
the statistical analysis of the ensemble of numerous cross-correlated
asset prices can reveal more reliably profitable trends in the market.
Typically, fund managers build diversified portfolios to decorrelate
constituent TF strategies as much as possible in order to enhance
their profit and reduce risk.  Relying on a Gaussian model with both
auto-correlation and cross-correlation structures of asset returns, it
was shown that conventional allocation schemes lead to sub-optimal
portfolios \cite{Grebenkov15}.  In particular, inter-asset
cross-correlations, if accounted for properly, can facilitate trend
detection and thus significantly improve the risk-adjusted portfolio
returns.  However, the optimal allocation of trend following
strategies developed in
\cite{Grebenkov15} remains too sophisticated for direct applications in
finance industry.  In particular, the optimal solution has to be
obtained by solving a very large system of nonlinear equations that
limits its implementation for large trading universes.  Moreover,
numerous parameters linked to the asset autocorrelation structures are
unknown and very difficult to estimate.

In the present paper, we extend the optimal allocation scheme
developed in \cite{Grebenkov15} in two directions.  On one hand, we
relax the former assumption of zero mean returns and include the
effect of small but always present net returns (risk premia).  While
their contribution is negligible at short time scales, mean returns
affect the allocation weights and thus the overall profitability of
the optimal portfolio at longer time horizons.  On the other hand, we
simplify as much as possible the covariance matrices accounting for
auto- and cross-correlations of assets.  Our goal here is to propose a
minimal theoretical setting that can produce explicit, easily
interpretable and practically implementable solutions of the
allocation problem.  In particular, we show that, under certain
assumptions specified below, the optimal solution can be seen as a
linear combination of four basic portfolios: risk parity, naive
Markowitz solution (when expectations of returns are based on trends),
trend on risk parity, and agnostic risk portfolio \cite{Benichou16}.
The choice of involving risk parity in the research design is
motivated by a more efficient way of allocating assets according to
their risk contribution to the portfolio because weights are
proportional to inverse volatility, which seeks more or less equal
risk-exposure between all the asset classes within a portfolio.  Risk
parity strategies are founded on the intuition of Black \cite{Black72}
that safer assets should offer higher risk-adjusted returns than
riskier assets \cite{Asness12}.  Moreover, a benefit over
mean-variance optimization, is that investors are not required to
formulate any assumption on the distribution of the returns
\cite{Chaves11}.

The paper is organized as follows.  In Sec. \ref{sec:theory}, we
introduce the autocorrelation model with the covariance matrix of
trends and risk premia.  We derive the main formula of the paper that
describes the optimal portfolio depending on the covariance matrix of
returns, the covariance matrix of trends and the risk premia.  In
Sec. \ref{sec:derivationofportfolios}, we introduce the specifications
for the covariance matrix of trends that make several basic portfolios
reported in the literature optimal from the theoretical point of view.
The optimal portfolio is obtained as a linear combination of these
basic portfolios.  In Sec. \ref{sec:backtest}, we present the
empirical backtest for different portfolios.  We compare them to the
simulated performance of their optimal linear combination.  Section
\ref{sec:conclusion} summarizes the main results, while technical
derivations are reported in Appendices.


\section{Mathematical model and its optimal solution}
\label{sec:theory}

We first extend the mathematical model of assets returns and linear
trend following strategies introduced in
\cite{Grebenkov15,Grebenkov14}.  We present then an approximate
optimal solution that maximizes the squared Sharpe ratio of the
portfolio.

\subsection{Mathematical model}
\label{sec:model}

First, we extend the mathematical model introduced in
\cite{Grebenkov15} by adding drift terms to describe the risk premia
that should be positive according to the theory.  We postulate that
the return%
\footnote{
Throughout this paper, 
we call by ``returns'' additive logarithmic returns resized by
realized volatility which is a common practice on futures markets
\cite{Martin12,Thomas12}.  Although asset returns are known to exhibit
various non-Gaussian features (so-called ``stylized facts''
\cite{Bouchaud,Mantegna,Mantegna95,Bouchaud01,Sornette03,Bouchaud04}),
resizing by realized volatility allows one to reduce, to some extent,
the impact of changes in volatility and its correlations
\cite{Bouchaud01b,Valeyre13, Valeyre19}, and to get closer to the Gaussian
hypothesis of returns \cite{Andersen00}.}
$r_t^j$ of the $j$-th asset at time $t$ has three contributions: a
constant drift $\mu^j$, an instantaneous fluctuation (noise)
$\ve_t^j$, and a stochastic trend, which is modeled as a linear
combination of random fluctuations $\xi^j_{t'}$,

\begin{equation}  \label{eq:rt0}
r_t^j = \mu^j + \ve_t^j + \sum\limits_{t'=1}^{t-1} \A^j_{t,t'} \xi_{t'}^j \,,
\end{equation}
where the matrix $\A^j$ describes the stochastic trend of the $j$-th
asset; in particular, when the elements of $\A^j$ decay exponentially
(see \ref{sec:derivation}), this is a discrete version of a stochastic
multi-asset price model in which the trends follow unobservable
correlated Ornstein-Uhlenbeck processes.  In turn, $\ve_1^j, \ldots,
\ve_t^j$ and $\xi_1^j, \ldots, \xi_t^j$ are two sets of independent
Gaussian variables with mean zero and the following covariance
structure:
\begin{equation}
\langle \ve_t^j \ve_{t'}^k\rangle = \delta_{t,t'} \C_\ve^{j,k}, \qquad
\langle \xi_t^j \xi_{t'}^k\rangle = \delta_{t,t'} \C_\xi^{j,k}, \qquad
\langle \ve_t^j \xi_{t'}^k\rangle = 0,
\end{equation}
where $\delta_{t,t'} = 1$ for $t = t'$ and $0$ otherwise, and $\langle
\ldots\rangle$ denotes the expectation.  Here $\C_\ve$ and $\C_\xi$
are the covariance matrices that describe inter-asset correlations of
noises $\ve_t^j$ and of stochastic trend components $\xi_t^j$,
respectively.  The covariance matrix of Gaussian asset returns is then
\begin{equation}
\label{eq:C}
\C_{t,t'}^{j,k} \equiv \langle r_t^j r_{t'}^k \rangle = \delta_{t,t'} \C_\ve^{j,k} +  \C_{\xi}^{j,k} (\A^j\A^{k,\T})_{t,t'} ,
\end{equation}
where $\T$ denotes the matrix transposition.  For each asset, the
stochastic trend induces auto-correlations due to a linear combination
of exogenous random variables $\xi_t^j$ which are independent from
short-time noises $\ve_t^j$.  Moreover, these {\it auto-correlations}
(described by the matrix $\A^j$) are considered to be independent from
{\it inter-asset cross-correlations} (described by matrices $\C_\ve$
and $\C_\xi$).  In particular, the covariance matrices $\C_\ve$ and
$\C_\xi$ do not depend on time.

A TF portfolio is composed of $n$ assets with positive or negative
weights $\Pi^j_t$, which are in general re-evaluated at each time $t$
(e.g., on daily basis).  Here $\Pi^j_t$ is the position%
\footnote{
The term ``position'' refers to the exposure or investment in a given
asset.  It is generally used in futures trading where position can be
either positive (long) or negative (short) \cite{Hull}.}
of the TF strategy on the $j$-th asset at time $t$, which is evaluated
as a weighted linear combination of the signals from all assets:
\begin{equation}
\Pi^j_t = \sum\limits_{k=1}^n \omega_t^{j,k} ~ s^k(r_1^k,\ldots, r_{t-1}^k) ,
\end{equation}
where $s^k(r_1^k,\ldots,r_{t-1}^k)$ is a signal based on past returns
of the $k$-th asset, with weights $\omega^{j,k}_t$ to be determined at
each time $t$.  The incremental profit-and-loss (P\&L) of a TF
portfolio (i.e., the total return of the portfolio at time $t$) is
\begin{equation}
\label{eq:dPNL}
\dPNL_t = \sum\limits_{j=1}^n r_t^j ~ \Pi^j_t = \sum\limits_{j,k=1}^n \omega^{j,k}_t ~r_t^j ~ s^k(r_1^k,\ldots, r_{t-1}^k) ,
\end{equation}
where $\omega^{j,k}_t$ can thus be interpreted as the weight of the
$k$-th signal onto the position of $j$-th asset.  The particular case
of diagonal weights (when $\omega^{j,k}_t = 0$ for $j\ne k$)
corresponds to a portfolio of $n$ TF strategies with weights
$\omega^{j,j}_t$.  Therefore, the standard portfolio allocation
problem is included in our framework, in which the diagonal weight
$\omega^{j,j}_t$ represents the amount of capital allocated to the
$j$-th asset.  In general, non-diagonal terms allow one to benefit
from inter-asset correlations to enhance the profitability of the TF
portfolio.

Following \cite{Grebenkov14}, we consider a TF strategy whose signal
is determined by a {\it linear} combination of earlier returns:
\begin{equation}
\label{eq:sk}
s^k(r_1^k,\ldots, r_{t-1}^k) = \sum\limits_{t'=1}^{t-1} \S^k_{t,t'} r_{t'}^k ,
\end{equation}
with given matrices $\S^k$.  In summary, the mathematical model is
fixed by choosing the vector $\mu$ of drifts $\mu^j$ and the matrices
$\C_\ve$, $\C_\xi$, $\A^j$, and $\S^k$.

\subsection{Optimal solution}
\label{sec:optimal}

Relying on the Gaussian character of the model, the mean, $\langle
\dPNL_t \rangle$, and the variance, $\var\{ \dPNL_t \}$, of
the incremental profit-and-loss $\dPNL_t$ can be computed
\cite{Grebenkov15}.  In \ref{sec:derivation}, we provide
general formulas for these two quantities for our extended model from
Sec. \ref{sec:model}.  Using these formulas, one can therefore search
for the weights $\omega^{j,k}_t$ that optimize a chosen criterion
(e.g., to minimize the variance under a fixed expected return for the
Markowitz theory).  In this paper, we aim at finding the optimal
weights $\omega^{j,k}_t$ that maximize the squared Sharpe ratio (or
squared risk-adjusted return of the portfolio),
\begin{equation}
\label{eq:Sharpe}
\Sh^2 \equiv \frac{\langle \dPNL_t \rangle^2}{\var\{ \dPNL_t \}} 
\end{equation}
(note that $\Sh^2$ is used instead of $\Sh$ just for convenient
notations, the optimization results are identical in both cases).  It
was shown in \cite{Grebenkov15} that this optimization problem is
equivalent to solving a set of $n^2$ quadratic equations onto $n^2$
unknown weights $\omega^{j,k}_t$ (see \ref{sec:derivation} for
details).  Since the mean and the variance of the increment P\&L
depend on time due to the dynamic character of TF strategies, the
optimal weights need to be re-evaluated at each time step of the TF
strategy.  Unfortunately, this formal solution is impractical due to
its computational costs for realistic trading universes with many
hundred of assets.  Moreover, the solution depends on numerous model
parameters (matrices $\C_\ve$, $\C_\xi$, $\A^j$ and $\S^k$) whose
accurate calibration from empirical data is not feasible.

These limitations motivated us to search for simplifications under
which practically relevant explicit solutions are possible.  The
fundamental challenge in forecasting next price moves follows from the
fact that the short-time noises $\ve_t^j$ provide the dominant
contributions to the returns.  In other words, the covariance matrix
of returns, $\C$, is essentially given by the matrix $\C_\ve$, whereas
the matrices $\C_\xi$ and $(\M)_{ij} = \mu^i \mu^j$ are negligible in
comparison to $\C_\ve$.  In this situation (and under some other, more
technical simplifications described in \ref{sec:derivation}), we
derive in \ref{sec:derivation} the explicit approximate expression for
the matrix $\omega$ of the optimal weights:
\begin{equation} \label{eq:omega_approx2a}
\omega_t \approx \C^{-1} \bigl(h_\xi(t) \C_\xi + h_\mu(t) \M\bigr) \C^{-1} ,
\end{equation}
where $h_\xi(t)$ and $h_\mu(t)$ are two explicitly known
time-dependent functions that determine relative contributions of
auto-correlation induced stochastic trends and net returns,
respectively.  At $t$ goes to infinity, functions $h_\xi(t)$ and
$h_\mu(t)$ tend to constants $h_\xi$ and $h_\mu$ that describe
relative contributions of stochastic trends and net returns in the
steady state regime:
\begin{equation} \label{eq:omega_approx2}
\omega \approx \C^{-1} \bigl(h_\xi \C_\xi + h_\mu \M\bigr) \C^{-1} .
\end{equation}
In this regime, the matrix $\omega$ of optimal weights does not depend
on time anymore.

This approximate optimal solution is the main theoretical result of
the paper.  At first thought, a linear superposition of two
contributions, stochastic trends (asset autocorrelations) and drifts,
is rather surprising given that, without our simplifications, one
would have to solve a large system of nonlinear equations.  The
linearity of the solution is a very appealing property.  In fact, it
simplifies the determination of the optimal portfolio, even though the
matrices $\M$ and $\C_\xi$, which are very difficult to estimate,
remain partly unknown.  Indeed, the optimal portfolio is a linear
combination of two basic portfolios, each of which can be determined
easier.  The first portfolio is based on the risk premia, $\M$, as if
stochastic trends (autocorrelation) did not exist.  The second
portfolio depends only on the covariance matrix of trends, $\C_\xi$,
as if the risk premia did not exist.

In the following, we focus on the steady state solution
(\ref{eq:omega_approx2}).  In the next section, we discuss
specifications for the matrices $\M$ and $\C_\xi$, under which the
solution leads to basic portfolios referred in the literature as
optimal.  We will then show that it is their linear combination that
is optimal.

\section{Derivation of basic particular and generalized optimal portfolios}
\label{sec:derivationofportfolios}

\subsection{Risk-parity portfolio (RP)}

We assume here that there is no autocorrelation in the returns (i.e.,
$\C_\xi = 0$) and that only drifts contribute (i.e., $\M \ne 0$).  The
approximate optimal solution is then
\begin{equation}  
\omega \approx h_\mu \, (\C^{-1} \mu)  (\mu^\dagger \C^{-1}) = h_\mu \, (\C^{-1} \mu)  (\C^{-1} \mu )^\dagger,
\end{equation}
where we used a direct product representation $\M = \mu \mu^\dagger$
with the vector $\mu$ of mean returns $\mu^j$.  Denoting by $\pi =
\C^{-1} \mu$ the vector of standard Markowitz weights, one gets
$\omega^{j,k} = h_\mu \pi^j \pi^k$.  As a consequence, the optimal
portfolio weights read
\begin{equation}
\Pi^j_t = \sum\limits_{k=1}^n \omega^{j,k} \, s^k = h_\mu \, \pi^j \, \biggl(\sum\limits_{k=1}^n \pi^k s^k\biggr).
\end{equation}
As the weights are determined up to an arbitrary multiplicative
factor, the sum in parentheses can be included into this constant,
yielding
\begin{equation}  \label{eq:Pi_RP}
\Pi_t \propto \pi = \C^{-1} \mu.
\end{equation}

To interpret this result as risk-parity portfolio, one can assume that
the Sharpe ratio is the same for all the instruments, i.e., the drift
$\mu^j$ is proportional to the volatility $\Sigma_{jj} =
\sqrt{\C_{jj}}$ of the asset $j$ (except for exchanges rates
instruments whose drift could be better assumed to be zero).  In that
way the risk is fairly rewarded and the drift describes a risk
premium.  This is the usual assumption made in the literature on the
``Maximum Diversification'' equity portfolio \cite{Choueifaty08}.
However, when it is applied to different asset classes (stocks, bonds,
commodities), the optimal portfolio is similar to the better known
risk parity portfolio that allows financial leverage and targets the
same risk on every asset classes, with the constraint to hold only
long positions.  This portfolio investment category, named ``risk
parity'', regroups massive investment.  For this reason, we use the
term ``risk parity portfolio'' instead of ``maximum diversification
portfolio''.  The weights of this portfolio can thus be written as
\begin{equation}  \label{eq:Pi_RP2}
\Pi_t \propto  \C^{-1} \Sigma.
\end{equation}
The interpretation of risk parity is rather simple: if the inter-asset
correlations could be neglected, i.e., $\C^{-1}$ would be diagonal,
and thus $\Pi_t \propto 1/\Sigma_{jj}$, i.e., it would be close to the
equally weighted (in volalility) portfolio.  It should be highly
correlated to the ``market mode'' of the correlation matrix, which is
most of the time the second eigenmode, when bonds and stocks are
negatively correlated.

Even though the assumption that average asset returns increase
proportionally with volatility is very approximative, it is a way to
get a proxy of the multi asset global market portfolio (the resulting
optimal portfolio will correspond to the market portfolio; the
equilibrium requires that all assets have a beta which is proportional
to its volatility divided by the volatility of the market portfolio).
We could have used a more complex assumption as for example in Ref.
\cite{Black92} but our assumption has the advantage to give an explicit
solution and one can expect that this optimal portfolio is a decent
proxy for the global market portfolio (see an attempt to measure the
inventory of a large universe of assets worldwide to proxy for a
theoretical market portfolio in \cite{Gadzinski18}; the portfolio was
called ``global market portfolio'' as it is composed of all risky
assets in the world in proportion to their market capitalization).
Note that the maximum diversification portfolios were found to have an
excess return quite similar to the capitalization weighted market
portfolio \cite{Clarke11}.  Therefore the risk parity portfolio
(\ref{eq:Pi_RP2}) captures the global average risk premium and plays a
very special role in asset management.

The drawback of this portfolio is its very high sensitivity to the
estimation of the correlation matrix: if cleaning of this matrix is
not good enough, some long-short positions, capturing fictitious
correlations, can appear.

\subsection{Naive Markowitz porfolio (NM)}

When expectations are based exclusively on trends, the optimal
Markowitz solution can be retrieved.  In fact, we consider here that
$\M = 0$, but conditional drifts can be represented through
stochatistic trends.  If we assume that $\C_\xi$ is proportional to
$\C$, then the naive Markowitz portfolio reads
\begin{equation}  \label{eq:Pi_NM}
\Pi \propto \C^{-1} \C_\xi \C^{-1} \s \propto \C^{-1} \s,
\end{equation}
where $\s$ is the vector of signals.  The portfolio is easy to
interpret as the result of independent trend-following strategies
applied to the eigenvectors $u_k$ of $\C$ with allocations in the
realized risk defined to be proportional to the inverse of the square
root of the eigenvalues $\lambda_k$ of $\C$.  In fact, using the
spectral decomposition of the matrix $\C$, one gets
\begin{equation}  \label{eq:Pi_NM2}
\Pi \propto \sum\limits_k \lambda_k^{-1} (u_k^\dagger \s)\, u_k ,
\end{equation}
i.e., the portfolio is a linear combination of eigenvectors $u_k$,
invested with the weights $\lambda_k^{-1} (\s^\dagger u_k)$.  Since
the return of a portfolio with weights given by an eigenvector $u_k$
has the variance $\var\{ \sum\nolimits_j u_k^j r^j \} = (u_k \C u_k) =
\lambda_k$, the factor $(\s^\dagger u_k)$ scales as $\lambda_k^{1/2}$
so that the above weights are proportional to $\lambda_k^{-1/2}$.

According to the naive Markowitz portfolio, the allocation would be
optimal if there were more trends on eigenvectors with small
eigenvalues that is not realistic from the financial point of view.

\subsection{Agnostic risk parity portfolio (ARP)}

To target the same unconditional risk on any eigenvector of the
correlation matrix, Benichou {\it et al.} proposed the agnostic risk
parity portfolio \cite{Benichou16}.  This specific asset allocation
allows to balance the risk between all the principal components of the
correlation matrix.  However, the optimality of the Sharpe ratio of
this portfolio was not discussed.  Here we suggest a simple sufficient
condition that makes the agnostic risk parity portfolio optimal.  We
assume that the correlation matrix of trends, $\tilde{\C}_\xi$, has
only one eigenvalue different from zero and that the associated
eigenvector is unknown.  This is coherent with the assumption of
Benichou {\it et al.} who considered the identity matrix to be the
best estimation of the correlation matrix of signals and not the
correlation matrix of returns.  In other words, the specification of
$\tilde{\C}_{\xi}$ means that trends are concentrated on only one risk
factor as if herding behavior could be efficient and amplified in only
one dimension in the each time moment (in \ref{sec:emergence}, we
introduce and discuss a very simple interaction model between agents
that generates such a pattern).  The trend on every instrument shares
therefore the same common but unknown factor, which is likely to
change from period to period and be for example the risk parity
factor, the fly-to-quality factor or more specific factor linked to a
local event as Brexit or specific initial trend as oil crash or
bubble.

Neglecting the matrix $\M$ of biases in returns, our optimal solution
(\ref{eq:omega_approx2}) implies
\begin{equation}
\Pi \approx h_\xi \, \C^{-1} \C_\xi \C^{-1} \s  ,
\end{equation}
where $h_\xi$ is a normalization constant.  The covariance matrices
$\C$ and $\C_\xi$ can be expressed in terms of the associated
correlation matrix $\tilde{\C}$ and the normalized covariance matrix
$\tilde{\C}_\xi$ as
\begin{equation}
\C = \Sigma \tilde{\C} \Sigma, \qquad \C_\xi = \Sigma \tilde{\C}_\xi \Sigma,
\end{equation}
where $\Sigma$ is again the diagonal matrix of volatilities:
$(\Sigma)_{ij} = \delta_{ij} \sqrt{(\C)_{ii}}$.  We denote by $\U$ the
matrix whose columns are composed of eigenvectors of the matrix
$\tilde{\C}_\xi$,
\begin{equation}
\tilde{\C}_\xi = \U \Nu \U^\dagger,
\end{equation}
where $\Nu$ is the diagonal matrix formed by the eigenvalues of
$\tilde{\C}_\xi$: $\nu_1,0,0,\ldots,0$.  Given the structure of the
matrix $\Nu$, it is convenient to split the matrix of eigenvectors as
\begin{equation}
\U = \U_{||} + \U_{\perp} ,
\end{equation}
where $\U_{||}$ contains only the eigenvector $u_1$ (corresponding to
$\nu_1$), while $\U_{\perp}$ contains the remaining eigenvectors.  The
optimal weights read then as
\begin{eqnarray}  \nonumber
\Pi &\approx& h_\xi \, \C^{-1} \Sigma (\U_{||} + \U_{\perp}) \Nu (\U_{||} + \U_{\perp})^\dagger \Sigma \C^{-1} \s  \\  
\label{eq:Pi_v}
&=& h_\xi \, \nu_1 \, (v_1^\dagger \, \tilde{\s}) \Sigma^{-1} v_1 ,
\end{eqnarray}
where $v_1 = \tilde{\C}^{-1} u_1$, $\tilde{\s} = \Sigma^{-1} \s$, and
we used the particular structure of the matrix $\Nu$.
Under the final assumption that $u_1$ is an $L_2$-normalized
eigenvector of the correlation matrix $\tilde{\C}$, one gets $v_1 =
\lambda_1^{-1} u_1$, where $\lambda_1$ is the corresponding
eigenvalue of $\tilde{\C}$.  The optimal weights are then
\begin{equation}  \label{eq:Pi_opt2}
\Pi \approx h_\xi \, \nu_1 \, \lambda_1^{-2} (u_1^\dagger\, \tilde{\s}) \Sigma^{-1} u_1 .
\end{equation}

We fix the normalization constant $h_\xi$ by requiring that the
variance of the portfolio,
\begin{equation}
V = (\Pi^\dagger \C \Pi) \approx h_\xi^2 \, \nu_1^2 \, \lambda_1^{-4} \, (u_1^\dagger \, \tilde{\s})^2 \, 
\underbrace{(u_1^\dagger \Sigma^{-1} \C \Sigma^{-1} u_1)}_{=\lambda_1}
\end{equation}
(here we neglected smaller contribution from $\C_\xi$), is equal to
$1$ {\it on average} over all directions of $u_1$:
\begin{equation}  \label{eq:Var_av}
1 = \langle V \rangle \approx h_\xi^2 \, \nu_1^2 \,  \lambda_1^{-3}  \, \langle(u_1^\dagger \, \tilde{\s})^2 \rangle ,
\end{equation}
We get then from Eq. (\ref{eq:Pi_opt2}):
\begin{equation}  \label{eq:Pi_opt0}
\Pi \approx  \frac{(u_1^\dagger \, \tilde{\s})}{\langle(u_1^\dagger \, \tilde{\s})^2 \rangle^{1/2}} 
\Sigma^{-1} \lambda_1^{-1/2} u_1 = \frac{(u_1^\dagger \, \tilde{\s})}{\langle(u_1^\dagger \, \tilde{\s})^2 \rangle^{1/2}} 
 \Sigma^{-1} \tilde{\C}^{-1/2} u_1 .
\end{equation}
The average of the portfolio over all possible (uniformly chosen)
directions $u_1$ yields
\begin{equation} \label{eq:Pi_ARP}
\langle \Pi \rangle \propto \Sigma^{-1} \tilde{\C}^{-1/2} \Sigma^{-1} \s ,
\end{equation}
where we omitted the proportionality constant $\langle(u_1^\dagger \,
\tilde{\s})^2 \rangle^{-1/2}$.  This is the agnostic risk parity
portfolio \cite{Benichou16}, which differs from the naive Markowitz
portfolio by the power $-1/2$ of the correlation matrix $\tilde{\C}$.

We note that the goal here was to provide a simple sufficient
condition under which the agnostic risk-parity portfolio would be
optimal.  To get Eq. (\ref{eq:Pi_ARP}) with the matrix
$\tilde{\C}^{-1/2}$, we employed numerous assumptions that make the
sufficient condition too restrictive.  We emphasize that the
sufficient condition is not the necessary one, and we expect the ARP
portfolio to be optimal under (much) weaker restrictions.

\subsection{Trend-on-risk-parity portfolio (ToRP)}

In this case, we assume that the special direction $u_1$ of the
normalized covariancen matrix of trends, $\tilde{\C}_\xi$, which was
supposed to be unknown in the agnostick risk parity portfolio, is
known and corresponds to the risk parity portfolio, $u_1 \propto {\bf
1}$, where ${\bf 1}$ is a vector composed of $1$ for all stocks except
for exchange rates instruments.  Earlier studies have not shown any
conclusive evidence for the direction of causality between interest
rates and stock prices for US markets \cite{Nozar88,Rahman97}.
Moreover, Chan {\it et al.}  questioned the existence of a common
trend between stock and bond prices \cite{Chan97}.  At the same time,
other works bring empirical evidences of a common part in stochastic
trends in international stock markets \cite{Kasa92} and explain why
the market mode can be an eigenvector of the matrix $\C_\xi$
\cite{Mellander92,Kim02,Lee12}.  These works justify our consideration
of the risk parity factor as the first eigenvector of the matrix
$\C_\xi$.
As we just saw, this direction plays a special role by capturing the
risk premium and thus helping to capture and to amplify the herding
behavior.
This means that trends on bonds and stocks are positively correlated,
and the best way to capture these trends is to measure the trend on
the risk parity portfolio.

Using Eq. (\ref{eq:Pi_v}) with $v_1 = \tilde{\C}^{-1} u_1$, we get
\begin{equation}   \label{eq:Pi_opt}
\Pi \propto \bigl((\Sigma {\bf 1})^\dagger \, \C^{-1} \s\bigr) \, \C^{-1} \Sigma {\bf 1} ,
\end{equation}
where the proportionality constant can be chosen by fixing the
variance of the portfolio, as in the agnostic risk-parity case.  Here,
we kept explicitly the scalar factor $\bigl((\Sigma {\bf 1})^\dagger \,
\C^{-1} \s\bigr)$, which depends on the signal $\s$.  This factor
corresponds to asset trends projected onto the risk parity portfolio
and thus represents the trend of risk parity portfolio.

The risk parity portfolio is a very particular portfolio as it
captures very well both the risk premia (or the carry) and a large
part of the trends.  Bhansali {\it et al.} confirm the link between
the carry and the trends as they show that the trend has a better
forecasting power when the carry is high \cite{Bhansali15}.  We can
also extent this section to the trend on other factors that can
capture the residual part of the trends.  As an example, the Value and
Momentum factors that could be good candidates are profitable in the
equity world but also in the cross asset world \cite{Asness13}.
Another possible extension would be to implement optimal equity market
neutral trend following strategies on factors listed in
\cite{McLean16,Valeyre}.  Hodges {\it et al. } show that trend on
factor is the most efficient way to make factor timing as the trend is
the best indicator to forecast the returns of the Value, Quality,
Momentum and the Low volality factors, among other indicators
including valuation, business cycle indicators \cite{Hodges17}.

\subsection{The optimal generalized portfolio}

In Sec. \ref{sec:optimal}, we have shown how the weights of the
optimal portfolio can be expressed via Eq. (\ref{eq:omega_approx2})
through the matrices $\C$, $\C_\xi$, and $\M$.  While the covariance
matrix $\C$ (or $\C_\ve$) can be estimated from empirical data, both
matrices $\C_\xi$ and $\M$ are very difficult to estimate.  In this
situation, it may be convenient to model the covariance matrix of
trends, $\C_\xi$ as a linear combination of the covariance matrices of
three basic portfolios discussed in this section: (i) the naive
Markowitz case ($\C_\xi \propto \C$); (ii) the agnostic risk parity
case ($\tilde{\C}_\xi$ has only one nonzero eigenvalue), and (iii) the
trend on the risk parity case ($\C_\xi$ has one eigenvector
corresponding to the risk parity).  Including also the matrix of net
returns, $\M$, the linearity of Eq. (\ref{eq:omega_approx2}) implies
that the optimal portfolio can be studied as a linear combination of
the four basic portfolios.  The empirical optimal weights could
therefore give a clue to estimate the covariance matrix of trends
$\C_\xi$.

\section{Empirical backtest}
\label{sec:backtest}
\subsection{Description of data and parameters}

We select the most liquid futures that include 24 futures on stock
index, 14 futures on bonds index and 9 futures on FOREX.  The period
starts from 8$^{\rm th}$ May 1985 and ends at 31$^{\rm st}$ December
2018.  The Sharpe ratio and backtest statistics are computed based on
the period from 1$^{\rm st}$ January 1993 to 27$^{\rm st}$ August
2020 (see Table \ref{tab:listofinsturment}).  We do not take into
account transactions cost and market impact.  In practice, other
constraints should be included to ensure the liquidity of the
portfolio and to minimize the market impact.  As we do not include
these constraints in the optimization, the implemented portfolio can
be different from the theoretical formula.

The signal of a TF strategy is chosen to be an EMA
\cite{Winters60,Brown}:
\begin{equation}
\label{eq:Sj}
\S_{t,t'} = \begin{cases}  (1-\eta)^{t-t'-1} , \quad t > t' , \cr 0 , \hskip 24mm t \leq t',  \end{cases}
\end{equation}
where $\eta$ is the rate of the TF strategy that we fix to be $\eta =
1/100$ on daily basis \cite{Grebenkov14}.  Setting the elements of
these matrices to $0$ for $t \leq t'$ implements the causality: the
signal at time $t$ relies only upon the earlier returns with $t' < t$.
Moreover, the same rate $\eta$ is used for all assets.

As mentioned in the footnote 1, it is convenient to consider the daily
returns resized by the realized volatility.  This resizing makes the
diagonal elements of the covariance matrix $\C$ to be very close to
$1$ so that $\C$ can be understood as the correlation matrix.
Although theoretical formulas in Sec. \ref{sec:derivationofportfolios}
were derived in the stationary regime (with a constant $\C$), it is
more practical to update the matrix $\C$ with time to render the
portfolio more reactive and sensitive to the latest changes in the
market.  For this reason, we estimate the matrix $\C$ as follows.
First, we estimate the covariance matrix $\hat{\C}$ of weekly returns
$\hat{r}_t^j$ to offset different trading hours used worldwide.  For
this purpose, we use an EMA with $\eta'=1/750$:
\begin{equation}
\hat{\C}_t^{ij} = (1-\eta') \hat{\C}_{t-1}^{ij} + \eta' \, \hat{r}^i_t \, \hat{r}^j_t .
\end{equation}
The covariance matrix is then rescaled by its diagonal elements:
\begin{equation}
\hat{\hat{\C}}_t^{ij} = \frac{\hat{\C}_t^{ij}}{\sqrt{\hat{\C}_t^{ii} \, \hat{\C}_t^{jj}}} \,.
\end{equation}
The latter is cleaned with the aid of the rotational invariant
estimator \cite{Bun16,Bun162} to finally get $\C$.

The variances $v_t^j$ are estimated from the daily returns.  In
fact, as the volatilities characterizes a single asset, the issue of
different trading hours is less relevant, and it is preferable to
estimate with more returns.  Here, we use an EMA with $\eta=1/100$
\begin{equation}
v^{j}_t = (1-\eta) v^{j}_{t-1} + \eta \, [r^j_t]^2.
\end{equation}
For the risk parity portfolio, we set $\mu_i =\sqrt{v^i_t}$ for stock
index and bonds, and $\mu_i = 0$ for exchange rates because exchange
rates present a long-short, completely neutral investment.

\begin{table}
\begin{center}
\begin{turn}{90}
\begin{tabular}{|c|c|c|} \hline
Stock indices & Bond indices & FOREX \\  \hline
\begin{tabular}[t]{l}
			AMSTERDAM EOE Idx\\
			S\&P Midcap 400 Idx e-mini\\
			Russell 2000 Idx e-mini\\
			Cac 40\\
			Dax\\
			Ftse 100\\
			STOXX Europe 600 Index Futures\\
			Hang Sen\\
			Mib S\&p-mif\\
			Nikkei 225 Osaka\\
			Topix\\
			Kospi 200\\
			Mini MSCI EAFE Index Future\\
			Ibex 35\\
			Mini MSCI Emerging Markets Index Future\\
			Idx-S\&P CNX Nifty\\
			Nasdaq E-mini\\
			MSCI Singapore Index Futures\\
			S\&P 500 e-mini\\
			MSCI Taiwan Index Futures\\
			Dj Euro Stoxx\\
			S\&P Canada 60-ME\\
			SPI 200 Idx\\
			Mini Dow Futures \\
\end{tabular} &
\begin{tabular}[t]{l}
			BUND 10Yr\\
			CAD Bond 10Yr\\
			Bobl\\
			Schatz\\
			Long-term Euro-btp\\
			Euro-buxl Futures\\
			LONG Gilt 10Yr\\
			10yr Fr Gov Bond\\
			US T-NOTE 5Yr\\
			JGB 10Yr\\
			US T-NOTE 10Yr\\
			US T-Note 2Yr\\
			Ultra T-Bonds Combined\\
			US T-BOND 30Yr\\
\end{tabular} &
\begin{tabular}[t]{l}

			AUD/USD Fut.\\
			GBP/USD Fut.\\
			CAD/USD Fut.\\
			EUR/USD Fut.\\
			US DOLLAR Idx\\
			JPY/USD Fut.\\
			MXN/USD Fut.\\
			NZD/USD Fut.\\
			CHF/USD Fut.\\
\end{tabular} \\  \hline
\end{tabular}
\end{turn}
\end{center}
\caption{
List of the 47 instruments. }
\label{tab:listofinsturment}
\end{table}

\subsection{Interpretation of the empirical results}

\begin{figure}
\begin{center}
\includegraphics[width=100mm]{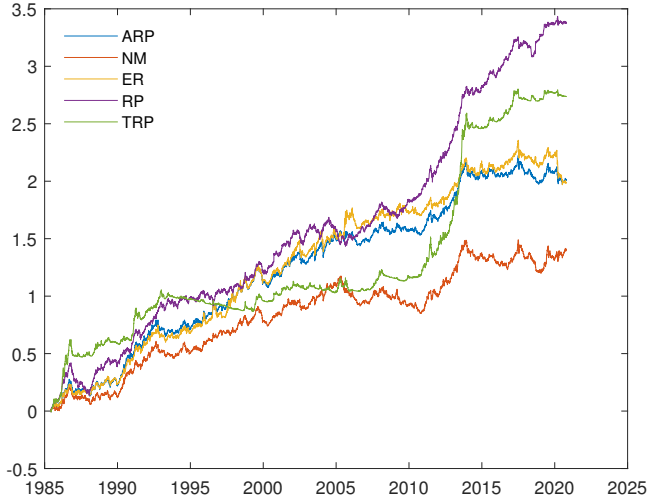}
\end{center}
\caption{
Equity curves of five portfolios: agnostic risk parity (ARP), naive
Markowitz (NM), equally weighted (EW), risk parity (RP), and trend
following on risk parity (ToRP).  Their Sharpe ratios are respectively
$0.75$, $0.52$, $0.65$, $1.24$, and $1.13$.  The Sharpe ratio of the
optimal combination with the risk parity portfolio is $1.37$, with the
optimal weights 19.5\% (ARP), 51\% (RP), and 30\% (ToRP).  The Sharpe
ratio of the optimal combination without the risk parity portfolio is
$1.18$, with
the optimal weights 28\% (ARP) and 72\% (ToRP). }
\label{fig:equity}
\end{figure}

In the practical implementation of the above portfolios, we adjust the
proportionality coefficient in Eqs. (\ref{eq:Pi_RP}, \ref{eq:Pi_NM},
\ref{eq:Pi_ARP}, \ref{eq:Pi_opt}) with time to target the same 
conditional volatility.  Figure \ref{fig:equity} shows the simulated
performance for the following portfolios: ARP (agnostic risk parity),
NM (naive Markowitz), EW (equally weighted), RP (risk parity), and
ToRP (Trend on Risk parity).

One can see that RP has the highest Sharpe ratio ($1.32$) as the risk
premia are significant and thus easier to capture as compared to
trends.  However, RP does not have appealing diversifying property
within the aggregated portfolio of all investors.  Among the trend
following portfolios that are decorrelated from RP (see Table
\ref{fig:correlation}), the ARP ($0.76$) performs much better than NM
($0.51$), suggesting that the assumption for the NM portfolio being
optimal is not realistic.  We see also that the ToRP ($1.19$) is the
best among the trend following ones, meaning that the common factor
for trends is most likely the risk parity portfolio.

\begin{figure}
\begin{center}
\includegraphics[width=100mm]{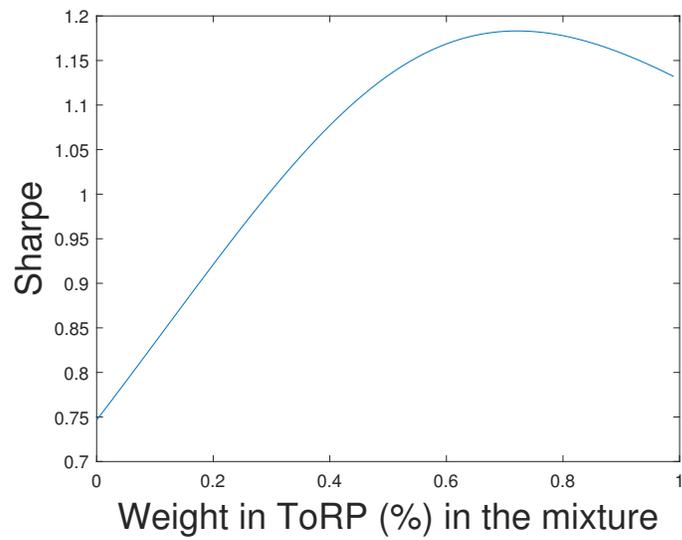}
\end{center}
\caption{
The Sharpe ratio of the combination between ARP and ToRP portfolios as
a function of the relative weight of ToRP.  The Sharpe ratio is above 1
for any mixture, in which the weight of ToRP is above 75\%. }
\label{fig:Sharpe}
\end{figure}

By excluding the RP that should be avoided to offer diversification to
investors, we also determined the optimal combination of ARP (27\%)
and ToRP (73\%) that improves the Sharpe ratio to $1.25$.  In
practice, ToRP has a shorter holding period (the Sharpe ratio is
expected to be smaller when including the market impact).  If one
allocates too much on the ToRP, it will reduce the capacity to manage
big assets of the portfolio, and increase the correlation between the
portfolio and RP.  Morever, the optimal weights are not so robust and
are very sensitive to the estimation of the Sharpe ratio of each
strategy that can change from period to period.  It is therefore more
robust to add a moderate contribution of ToRP to ARP to improve the
Sharpe ratio of ARP.  The weights of ARP at 75\% and of ToRP at 25\%
appear to be a good compromise.  The Sharpe ratio of the mixture
remains above 1 (Fig. \ref{fig:Sharpe}).

\begin{figure}
\begin{center}
\includegraphics[width=100mm]{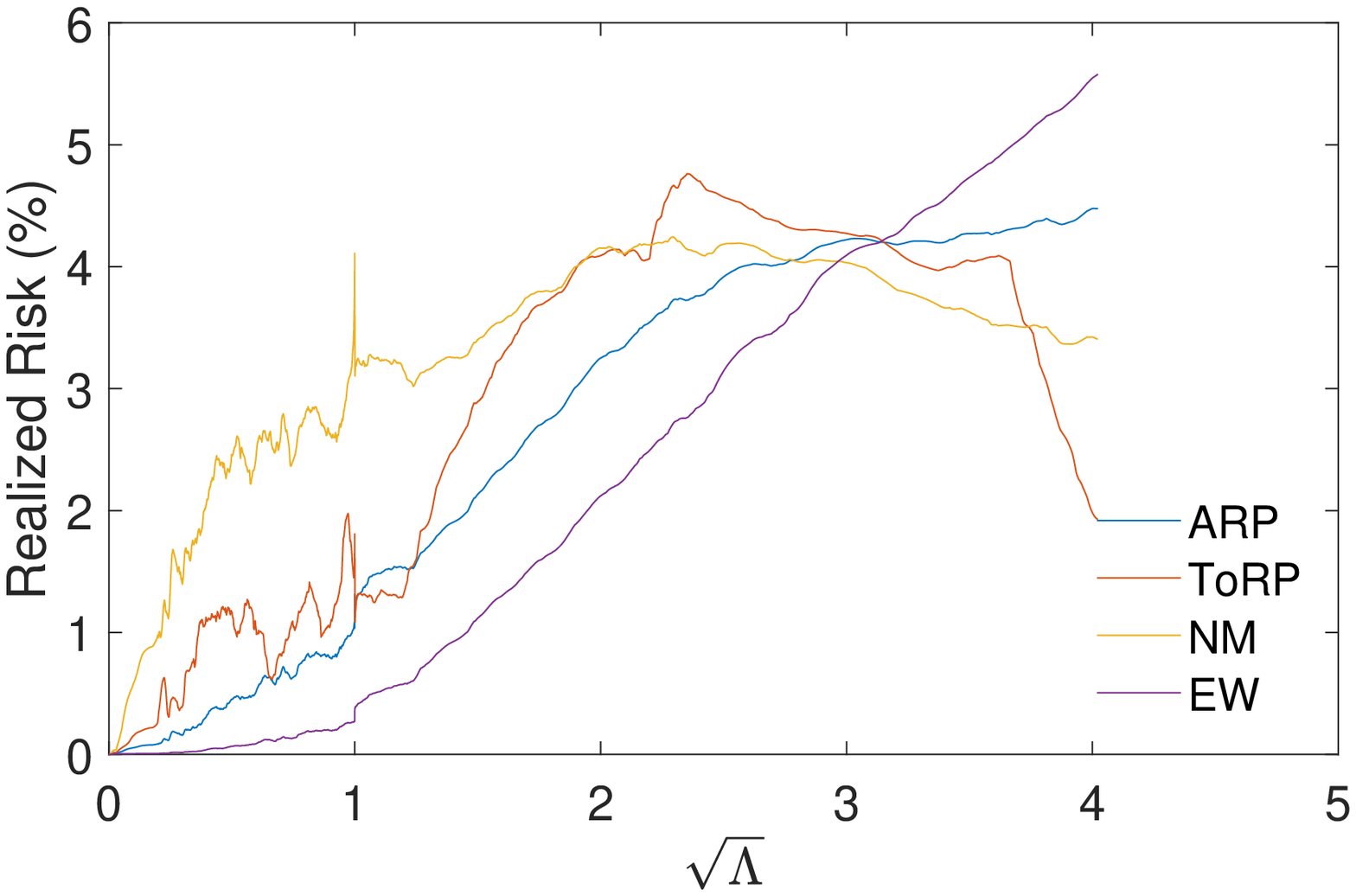}
\end{center}
\caption{
Empirical realized risk as a function of the square root of the
eigenvalue of the correlation matrix $\tilde{\C}$ of four portfolios:
agnostic risk parity (ARP), naive Markowitz (NM), equally weighted
(EW), trend following on risk parity (ToRP).  Disagreements with the
theoretically expected realized risk are observed (for example, the
risk is expected to be constant for ARP, inversely proportional to
$\sqrt{\Lambda}$ for NM, and proportional to $\sqrt{\Lambda}$ for EW).
Note that the NM is more allocated on small eigenvalues, while ToRP is
invested on intermediate eigenvalues (mainly the second one). }
\label{fig:realizedRisk}
\end{figure}

The EW portfolio works pretty well by two reasons: first, the universe
is well equilibrated between the number of stocks indices and the
number of bond indices; second, the EW portfolio, whose realized risk
is in theory proportional to the square root of the eigenvalue, could
be interpreted as a combination between ARP (the same realized risk
for any eigenmode, in theory) and ToRP (concentration of realized risk
in the first eigenmodes).  This interpretation is confirmed by
Fig. \ref{fig:realizedRisk} that shows empirical realized risk
depending on the square root of the eigenvalues.  The disagreement
with the backtest could be explained by the challenge to measure
properly realized risk on small eigenvalues and by the deviation
between the model and the market (in particular, correlations are not
constant in time).

\begin{table}[ht]
\centering
\begin{tabular}{|c|c|c|c|}  \hline
     & ARP  & RP   & ToRP \\   \hline \rule{0pt}{4ex}
 ARP &      & 0.23 & 0.34 \\   \hline \rule{0pt}{4ex}
  RP & 0.23 &      & 0.59 \\   \hline \rule{0pt}{4ex}
ToRP & 0.34 & 0.59 &      \\  \hline
\end{tabular}
\caption{
Correlations between the returns of three best portfolios: agnostic
risk parity (ARP), risk parity (RP), and trend following on risk
parity (ToRP).  The portfolio that is the most decorrelated to RP is
ARP, whereas ToRP is more correlated because it detected more often
buy signals than sell signals in the past.}
\label{fig:correlation}
\end{table}

\section{Conclusion}
\label{sec:conclusion}

We derive a theoretical setting to yield implementable solutions of
the allocation problem of trend following portfolios.  The main
formula of the paper describes the optimal portfolio as depending on
the covariance matrix of returns, the covariance matrix of trends and
the risk premia.

We implement the formula to gauge the performance of five well
established portfolios (Agnostic Risk Parity, Markowitz, Equally
Weighted, Risk Parity and Trend on Risk Parity), using daily data from
futures markets of 24 stock indexes, 14 bonds indexes and 9 FX, from
1985 to 2020.

Our main empirical finding is the optimal combination of the three
best portfolios produces a Sharpe ratio of $1.37$, with their respective
optimal weights of 19.5\% (ARP), 51\% (RP), and 30\% (ToRP)  which combines both traditional and alternative approach.  Consistent
with related recent literature, we confirm that RP portfolio, which is a proxy of the traditional and well diversified portfolio is a
important driver of performance.  Furthermore, we show
that the combination between ARP and ToRP is the best solution in term
of Sharpe ratio for the trend following approach and the alternative benchmark as they tend to minimize the correlation among assets.

\vskip 5mm




\appendix

\section{Derivation of the main results}
\label{sec:derivation}

In \cite{Grebenkov15}, we considered the model without drifts, $\mu^j
= 0$, for which the mean and the variance of the incremental
profit-and-loss $\dPNL_t^{(0)}$ were derived
\begin{equation}
\label{eq:mean_var}
\begin{split}
\langle \dPNL_t^{(0)} \rangle & = \sum\limits_{j,k=1}^n  \omega^{j,k}_t ~M^{j,k(0)}_t,  \\
\var\{ \dPNL_t^{(0)} \} & = \sum\limits_{j_1,k_1,j_2,k_2=1}^n  \omega^{j_1,k_1}_t \omega^{j_2,k_2}_t V^{j_1,k_1;j_2,k_2(0)}_t, \\
\end{split}
\end{equation}
where the superscript $0$ highlights the driftless character, and
\begin{eqnarray}
\label{eq:Mt_general}
\mathcal{M}^{j,k(0)}_t &=& \C_{\xi}^{j,k} (\S^k \A^k\A^{j,\T})_{t,t} , \\
\nonumber
V^{j_1,k_1; j_2,k_2(0)}_t & =& \C_\ve^{j_1,j_2} \C_\ve^{k_1,k_2} (\S^{k_1} \S^{k_2,\T})_{t,t} + 
 \C_\ve^{j_1,j_2} \C_{\xi}^{k_1,k_2} (\S^{k_1} \A^{k_1} \A^{k_2,\T} \S^{k_2,\T})_{t,t} \\
\nonumber
&+& \C_\ve^{k_1,k_2}  \C_\xi^{j_1,j_2} (\S^{k_1} \S^{k_2,\T})_{t,t} (\A^{j_1}\A^{j_2,\T})_{t,t} \\
\nonumber
&+&  \C_\xi^{j_1,j_2} (\A^{j_1}\A^{j_2,\T})_{t,t} \C_{\xi}^{k_1,k_2} 
(\S^{k_1} \A^{k_1} \A^{k_2,\T} \S^{k_2,\T})_{t,t}   \\
\label{eq:Vt_general}
&+& \C_{\xi}^{j_1,k_2} \C_{\xi}^{k_1,j_2} (\S^{k_1} \A^{k_1,\T} \A^{j_2})_{t,t}  
(\S^{k_2} \A^{j_1,\T} \A^{k_2})_{t,t} ,
\end{eqnarray}
where the matrices $\C_\ve$, $\C_\xi$, $\A^j$, and $\S^j$ are defined
in Sec. \ref{sec:model}.  The structural separation between
auto-correlations and inter-asset cross-corrections from
Eq. (\ref{eq:C}) is also reflected in these formulas.

Now we relax the former assumption of zero mean returns by adding
constant drifts $\mu^j$.  Employing the standard tools for averaging
Gaussian variables, one can evaluate the mean and variance of this
P\&L.  First, we get
\begin{equation}
\langle \delta \P_t \rangle = \langle \delta \P_t^{(0)} \rangle + 
\sum\limits_{j,k=1}^n \omega^{j,k}_t \sum\limits_{t'=1}^{t-1} \S_{t,t'}^k \, \mu^j \, \mu^k .
\end{equation}
Denoting
\begin{equation}  \label{eq:Shat}
\hat{S}^k_t = \sum\limits_{t'=1}^{t-1} \S_{t,t'}^k ,
\end{equation}
one has
\begin{equation}
\langle \delta \P_t \rangle = \sum\limits_{j,k=1}^n \omega^{j,k}_t \, \mathcal{M}_t^{j,k} ,
\end{equation}
with
\begin{equation}
\mathcal{M}_t^{j,k} = \mathcal{M}_t^{j,k(0)} + \mu^j \mu^k \hat{S}_t^k ,
\end{equation}
in which $\mathcal{M}_t^{j,k(0)}$ is given by Eq. (\ref{eq:Mt_general}).

Similarly, long but straightforward computations yield
\begin{equation}  \label{eq:varP}
\var\{ \delta \P_t \} = \sum\limits_{j_1,k_1,j_2,k_2=1}^n \omega^{j_1,k_1}_t \, \omega^{j_2,k_2}_t \, V_t^{j_1,k_1;j_2,k_2} ,
\end{equation}
with
\begin{align} \nonumber
V_t^{j_1,k_1;j_2,k_2} & = V_t^{j_1,k_1;j_2,k_2(0)} 
+ \mu^{j_1} \mu^{j_2} \bigl(\C_\ve^{k_1,k_2} [\S^{k_1} \S^{k_2,\dagger} ]_{t,t} + \C_\xi^{k_1,k_2} 
[\S^{k_1} \A^{k_1} \A^{k_2,\dagger} \S^{k_2,\dagger} ]_{t,t}\bigr)  \\
\nonumber
& + \mu^{j_1} \mu^{k_2} \hat{S}^{k_2}_t \C_\xi^{k_1,j_2} [\S^{k_1} \A^{k_1} \A^{j_2,\dagger} ]_{t,t} 
+ \mu^{j_2} \mu^{k_1} \hat{S}^{k_1}_t \C_\xi^{k_2,j_1} [\S^{k_2} \A^{k_2} \A^{j_1,\dagger} ]_{t,t} \\
\label{eq:Vt_new}
& + \mu^{k_1} \mu^{k_2} \hat{S}^{k_1}_t \hat{S}^{k_2}_t \bigl(\C_\ve^{j_1,j_2} + \C_\xi^{j_1,j_2} [\A^{j_1} \A^{j_2,\dagger} ]_{t,t}\bigr),
\end{align}
in which $V_t^{j_1,k_1;j_2,k_2(0)}$ is given by
Eq. (\ref{eq:Vt_general}).

Once the mean and the variance of the incremental P\&L are known, the
dynamic allocation problem for a portfolio of trend following
strategies is reduced to the standard optimization problem for a
portfolio composed of $n^2$ ``virtual'' assets (indexed by a double
index $j,k$) whose means are $\mathcal{M}^{j,k}_t$ and the covariance is
$V^{j_1,k_1;j_2,k_2}_t$.  One can therefore search for the weights
$\omega^{j,k}_t$ that optimize a chosen criterion (e.g., to minimize
the variance under a fixed expected return for the Markowitz theory).
Here we aim to maximize the squared Sharpe ratio (or squared
risk-adjusted return of the portfolio) in Eq. (\ref{eq:Sharpe}) that
reads
\begin{equation}
\label{eq:Sharpe10}
\Sh^2 = \frac{(\mathcal{M}_t^\T \omega_t)^2}{(\omega_t^\T V_t \omega_t)} .
\end{equation}
The optimization leads to the following equations on the weights
$\omega^{j,k}_t$:
\begin{equation}
\frac{\partial \Sh^2}{\partial \omega^{j,k}_t} = \frac{2(\mathcal{M}_t^\T \omega_t)}{(\omega_t^\T V_t \omega_t)^2} 
\bigl[ \mathcal{M}_t^{j,k}  (\omega_t^\T V_t \omega_t) - (V_t \omega_t)^{j,k} (\mathcal{M}_t^\T \omega_t)\bigr] = 0   \qquad (j,k =1,\ldots, n),
\end{equation}
or, equivalently,
\begin{equation}
\label{eq:optim_equations}
\sum\limits_{j_1,k_1,j_2,k_2=1}^n \bigl[\mathcal{M}_t^{j,k} V_t^{j_1,k_1;j_2,k_2} - V_t^{j,k;j_1,k_1} \mathcal{M}_t^{j_2,k_2} \bigr]
\omega^{j_1,k_1}_t \omega^{j_2,k_2}_t = 0 
\end{equation}
for all indices $j,k=1,\ldots,n$.  This is a set of $n^2$ quadratic
equations onto $n^2$ unknown weights $\omega^{j,k}_t$.  Since $\mathcal{M}_t$
and $V_t$ depend on time due to the dynamic character of TF
strategies, the optimal weights need to be re-evaluated at each time
step of the TF strategy.

\subsection{Approximate solution of the general problem}

The optimal weights $\omega^{j,k}_t$ satisfy
Eqs. (\ref{eq:optim_equations}), which can be re-written with $\mathcal{M}_t$
and $V_t$ as
\begin{eqnarray}  \label{eq:auxil22}
\mathcal{M}_t^{j,k} \biggl(\sum\limits_{j_1,k_1,j_2,k_2=1}^n \omega_{j_1,k_1} V_t^{j_1,k_1;j_2,k_2} \omega^{j_2,k_2}_t\biggr)
&=& \sum\limits_{j_1,k_1=1}^n V_t^{j,k;j_1,k_1} \omega^{j_1,k_1}_t \\
\nonumber
&\times& \biggl(\sum\limits_{j_2,k_2=1}^n \mathcal{M}_t^{j_2,k_2} \omega^{j_2,k_2}_t\biggr) 
\end{eqnarray}
for all $j,k = 1,\ldots,n$.  Treating the two sums in parentheses as
(unknown) constants and thinking of $\omega^{j,k}_t$ and $\mathcal{M}_t^{j,k}$
as vectors (with a double index $j,k$), one might wish writing an
explicit solution in the form
\begin{equation}  \label{eq:omega_formal}
\omega_t = c V_t^{-1} \mathcal{M}_t ,
\end{equation}
where $c$ is an arbibrary normalization constant, and $V_t^{-1}$ is
the ``inverse'' of $V_t$.  Given the sophisticated tensorial structure
of $V_t$ in Eqs. (\ref{eq:Vt_general}, \ref{eq:Vt_new}), the
definition of its inverse and thus the meaning of
Eq. (\ref{eq:omega_formal}) are problematic in general.

Here we discuss two assumptions under which such an explicit solution
is possible.  First, we assume that the matrices $\S^j$ and $\A^j$
describing the signal and the autocorrelation structure of asset
returns are the same for all stocks, i.e., $\S^j = \S$ and $\A^j =
\A$.  The sum over $j_1$ and $k_1$ in the right-hand side of
Eq. (\ref{eq:auxil22}) can be understood as a matrix (with respect to
indices $j$ and $k$) and shortly denoted as $[V_t \omega_t]_{j,k}$.
According to Eqs. (\ref{eq:Vt_general}, \ref{eq:Vt_new}), this matrix
can be written as
\begin{eqnarray}  \label{eq:Vt_matrix}
&& V_t \omega_t = f_{\ve\ve}(t) \C_\ve \omega_t \C_\ve +
f_{\ve\xi}(t) \C_\ve \omega_t \C_\xi + f_{\xi\ve}(t) \C_\xi \omega_t \C_\ve \\ \nonumber 
&& + \C_\xi \bigl(f_{\xi\xi}^{(1)}(t) \omega_t +
f_{\xi\xi}^{(2)}(t) \omega_t^\dagger \bigr)\C_\xi + \M \omega_t
(f_{\ve\ve}(t) \C_\ve + f_{\ve\xi}(t) \C_\xi) \\ \nonumber 
&& + [\hat{S}_t]^2 \bigl(\C_\ve + \C_\xi [\A\A^\dagger]_{t,t} \bigr) \omega_t
\M + \hat{S}_t (\S\A\A^\dagger)_{t,t} \bigl( \M \omega_t^\dagger \C_\xi
+ \C_\xi \omega_t^\dagger \M\bigr) ,
\end{eqnarray}
where $\M$ is the matrix of drifts, $\M_{j,k} = \mu^j \mu^k$, and
\begin{subequations}  \label{eq:f_def}
\begin{eqnarray}
f_{\ve\ve}(t) &=& (\S \S^\dagger)_{t,t} , \\
f_{\ve\xi}(t) &=& (\S \A \A^\dagger \S^\dagger)_{t,t} , \\
f_{\xi\ve}(t) &=& (\S \S^\dagger)_{t,t} (\A \A^\dagger)_{t,t} , \\
f_{\xi\xi}^{(1)}(t) &=& (\S \A \A^\dagger \S^\dagger)_{t,t} (\A \A^\dagger)_{t,t} , \\
f_{\xi\xi}^{(2)}(t) &=& [(\S \A^\dagger \A)_{t,t}]^2  .
\end{eqnarray}
\end{subequations}

Second, we assume that autocorrelations are weak so that one can
neglect terms which are of the second order in the matrix $\C_\xi$.
Similarly, we neglect terms containing both $\M$ and $\C_\xi$ as
drifts are as well small.  In this case, the above expression can be
approximated as
\begin{equation}
V_t \omega_t \approx f_{\ve\ve}(t)  \bigl(\C_\ve + g_1(t) \C_\xi + \M \bigr) \omega_t \bigl(\C_\ve + g_2(t) \C_\xi + g_3(t) \M \bigr) ,
\end{equation}
where
\begin{equation}
g_1(t) = \frac{f_{\xi\ve}(t)}{f_{\ve\ve}(t)} , \qquad 
g_2(t) = \frac{f_{\ve\xi}(t)}{f_{\ve\ve}(t)} , \qquad
g_3(t) = \frac{[\hat{S}_t]^2}{f_{\ve\ve}(t)} .
\end{equation}
Rewriting Eqs. (\ref{eq:auxil22}) in a matrix form as $V_t
\omega_t = c \mathcal{M}_t$ with an unknown constant $c$, one can finally
invert this matrix relation to get
\begin{equation}
\omega_t \approx \bigl(\C_\ve + g_1(t) \C_\xi + \M \bigr)^{-1} \bigl( h_\xi(t) \C_\xi + h_\mu(t) \M \bigr)
 \bigl(\C_\ve + g_2(t) \C_\xi + g_3(t)\M \bigr)^{-1} ,
\end{equation}
with
\begin{equation}
h_\xi(t) = c \, \frac{(\S \A \A^\T)_{t,t}}{f_{\ve\ve}(t)} \,,  \qquad 
h_\mu(t) = c \, \frac{\hat{S}_t}{f_{\ve\ve}(t)} \,.
\end{equation}
Here, the unknown constant $c$ is included into functions $h_\xi(t)$
and $h_\mu(t)$.  This is an approximate optimal solution for the
matrix of weights $\omega^{j,k}_t$.  Its explicit, easily computable
matrix form is one of the main theoretical results of the paper.  In
this solution, the matrices $\A$ and $\S$ determining assets
auto-correlation and TF signals, induce time dependence via the
functions $g_1(t)$, $g_2(t)$, $g_3(t)$, $h_\xi(t)$ and $h_\mu(t)$.  We
emphasize that the impact of time dependence is in general highly
nontrivial given that the functions $g_i(t)$ stand in front of
matrices in a linear combination which is inverted.

Neglecting again the contribution of small matrices $\C_\xi$ and $\M$
(as compared to $\C_\ve$), we get a practical approximation of the
optimal solution:
\begin{equation} \label{eq:omega_approx20}
\omega_t \approx \C_\ve^{-1} \bigl(h_\xi(t) \C_\xi + h_\mu(t) \M\bigr) \C_\ve^{-1} .
\end{equation}
In this approximation, the impact of time dependence is explicit:
functions $h_\xi(t)$ and $h_\mu(t)$ determine relative contributions
of auto-correlation induced stochastic trends and net returns,
respectively.  We recall that these functions are determined up to an
arbitrary multiplicative factor so that an additional constraint on
the optimal portfolio will be needed to fix the weights (e.g., the
targeted variance of the portfolio).  In the stationary regime, the
functions $h_\xi(t)$ and $h_\mu(t)$ reach their limits, denoted
$h_\xi$ and $h_\mu$.  Finally, the covariance matrix of instantaneous
fluctuations of returns, $\C_\ve$, is close, in the leading order, to
the covariance matrix of returns, $\C$.  We can thus replace $\C_\ve$
by $\C$ to rewrite Eq. (\ref{eq:omega_approx20}) in the form
(\ref{eq:omega_approx2a}) presented in the text.

\section{Emergence of the dominant factor: an interacting agents model}
\label{sec:emergence}

In this Appendix, we discuss a simple model of interacting agents to
rationalize the emergence of the dominant factor.  This model is
inspired by studies of collective opinions shifts and other models of
statistical physics \cite{Michard05,Bouchaud13}.  We emphasize that
this model is fully unrelated to our model from Sec. \ref{sec:theory}
and serves exclusively to provide complementary support to empirical
evidences of the dominant factor.

We suppose that there are $A$ interacting trading agents.  At each
moment of time $t$, each agent adopts one of $N$ available trading
strategies (that could correspond to $N$ portfolios based on $N$
eigenvectors of the covariance matrix).  To describe this choice, we
introduce a matrix $S(t)$ of size $A\times N$ whose element
$S_{a,n}(t)$ is equal to $1$ if the agent $a$ adopts the $n$-th
strategy at time $t$, and $0$ otherwise: $S_{a,k}(t) = \delta_{k,n}$.
At the next time step $t+1$, each agent re-evaluates his strategy in
the following way: first, one computes the ``preference matrix''
$X_{a,k}$ of the $a$-th agent to the strategy $k$,
\begin{equation}
X_{a,k} = \ve_{a,k} + \sum\limits_{a'=1}^A J_{a,a'} S_{a',k}(t),
\end{equation}
where $\ve_{a,k}$ is the individual $a$-th agent's preference to the
strategy $k$, while the matrix $J_{a,a'}$ characterizes to which
extent the agent $a$ is influenced by another agent $a'$; second, for
each agent $a$, one selects the strategy $k_{\rm max}$ with the
maximal preference $X_{a,k_{\rm max}}$ among all $X_{a,1}, X_{a,2},
\ldots, X_{a,N}$, and sets $S_{a,k}(t+1) = \delta_{k,k_{\rm max}}$.
In other words, the agent $a$ adopts for time $t+1$ the strategy that
was most preferred for him at time $t$.  If there was no interaction
with other agents (i.e., $J_{a,a'} = 0$ for all $a,a'$), each agent
would keep its preferred strategy that corresponds to the maximum of
$\ve_{a,1}, \ve_{a,2}, \ldots, \ve_{a,N}$.  In the presence of
interactions, the agent selects his strategy as a compromise between
his own individual preferences (characterized by $\ve_{a,k}$) and the
influence of other agents and their preferred strategies.  In the
ultimate limit when the interactions are all equal and very high,
$J_{a,a'} = J \to \infty$, if there is a single strategy adopted by
the largest number of agents at the beginning, then this strategy will
provide the maximum of $X_{a,k}$ for all agents and thus will be
adopted by all agents at the next step.  Clearly, one can expect a
transition between the no interaction limit (when each agent keeps
using its preferred strategy) and the strong interaction limit (when
all agents use the same strategy).

The emergence of the dominant mode (i.e., a single strategy adopted by
all agents) depends on the matrices $J_{a,a'}$ and $\ve_{a,k}$
governing the dynamics.  In statistical physics, it is common that
fine details of the model parameters do not matter in the limit of a
large number of particles (here, agents).  The same kind of
universality is expected for the present model.  We perform
simulations to illustrate that the overall amplitude of interactions
(as compared to individual preferences $\ve_{a,k}$) is the major
parameter that determines the transition.  

Without dwelling on the analysis of this model, we make a simple
choice of the parameters: the interaction matrix $J_{a,a'}$ is
considered to be constant, $J_{a,a'} = J$, i.e., all agents have the
same level of influence on each other.  In turn, the elements of the
matrix $\ve$ are independent centered normally distributed numbers
with unit variance.  The initial state $S_{a,k}(0)$ is also set
randomly, by selecting for each agent $a$ one preferred strategy among
$N$ available with a uniform law.  In this setting, the level of
interactions $J$ (as compared to the unit level of individual
preferences) is the major parameter, along with the number of agents
$A$ and the number of strategies $N$.  Each simulation is performed
for $T$ steps, with $T$ being chosen to allow for convergence to a
steady state, resulting in the matrix $S_{a,k}(t)$ at all time steps
$t=0,1,2,\ldots,T$.  From this basic quantity, we compute the
empirical average over all agents,
\begin{equation}
\overline{S}_k(t) = \frac{1}{A} \sum\limits_{a=1}^A S_{a,k}(t) ,
\end{equation}
which represents the overall interest of agents into the $k$-th
strategy at time $t$.  Given that all $\overline{S}_k(t)$ are still
random variables, we repeat simulations $M$ times to approximate the
expectation $I_k(t) = \E\{ \overline{S}_k(t)\}$ by averaging out
random fluctuations among simulated results.

By definition, each $I_k(t)$ is a number from $0$ to $1$ such that
their sum over $k$ is equal to $1$.  In other words, $I_k(t)$ can be
interpreted as the average fraction of agents interested in the $k$
strategy.  At the beginning, the uniform assignment of preferred
strategies among the agents yields $I_k(0) = 1/N$, i.e., all
strategies are equally preferred.  As time goes on, interactions
between agents can spontaneously break the initial symmetry between
all strategies and lead to the emergence of a dominant strategy
preferred by the majority of agents.  In the following, we will
illustrate the behavior of $I_k(t)$ for different choices of the
parameters.  We will also look at the dynamics of the largest
fraction, i.e., how $\max_k \{I_k(t)\}$ evolves with time.  In
particular, we will see how the steady-state value of this maximum
depends on the level of interactions $J$.

\begin{figure}
\begin{center}
\includegraphics[width=65mm]{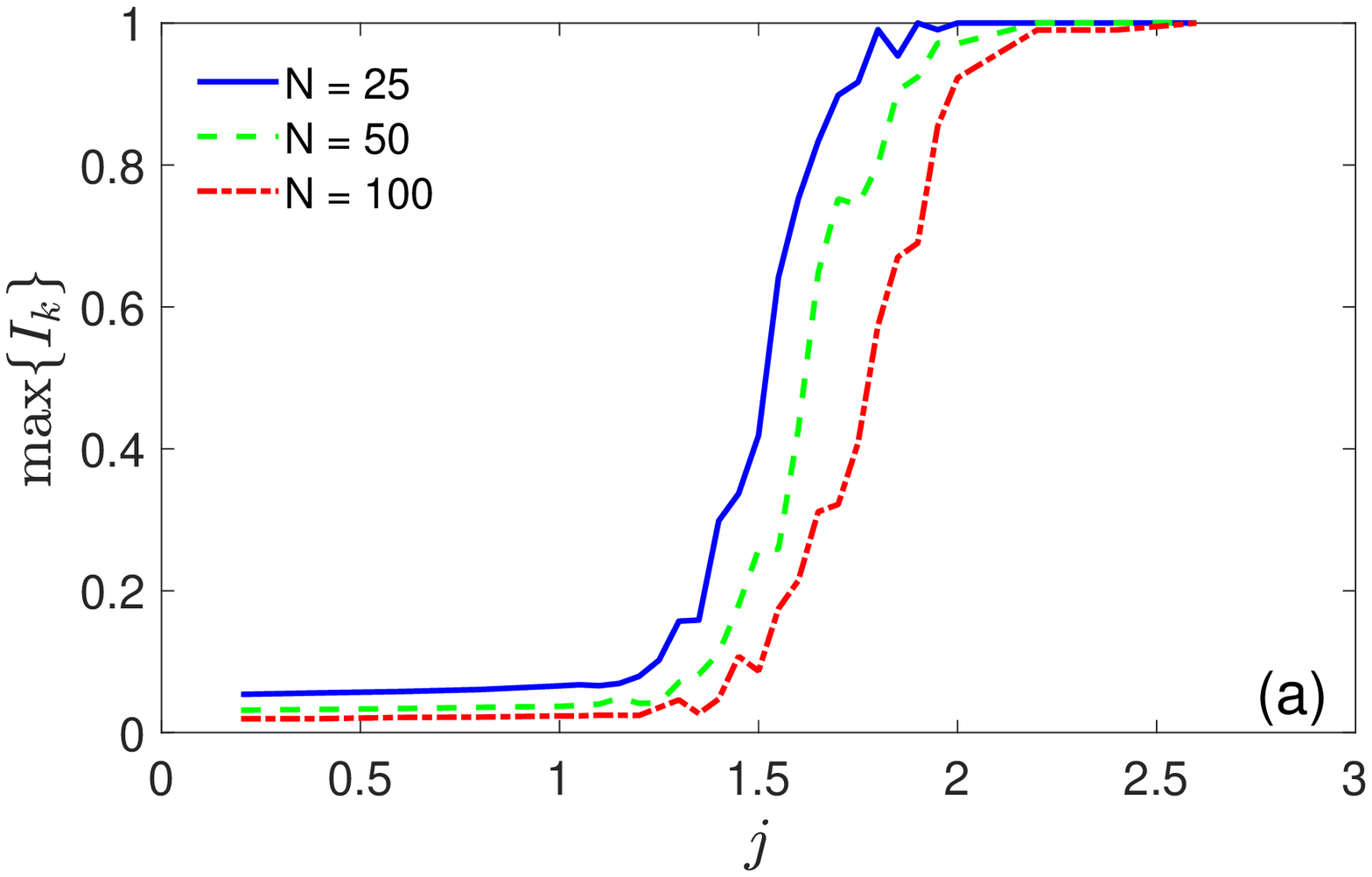}
\includegraphics[width=65mm]{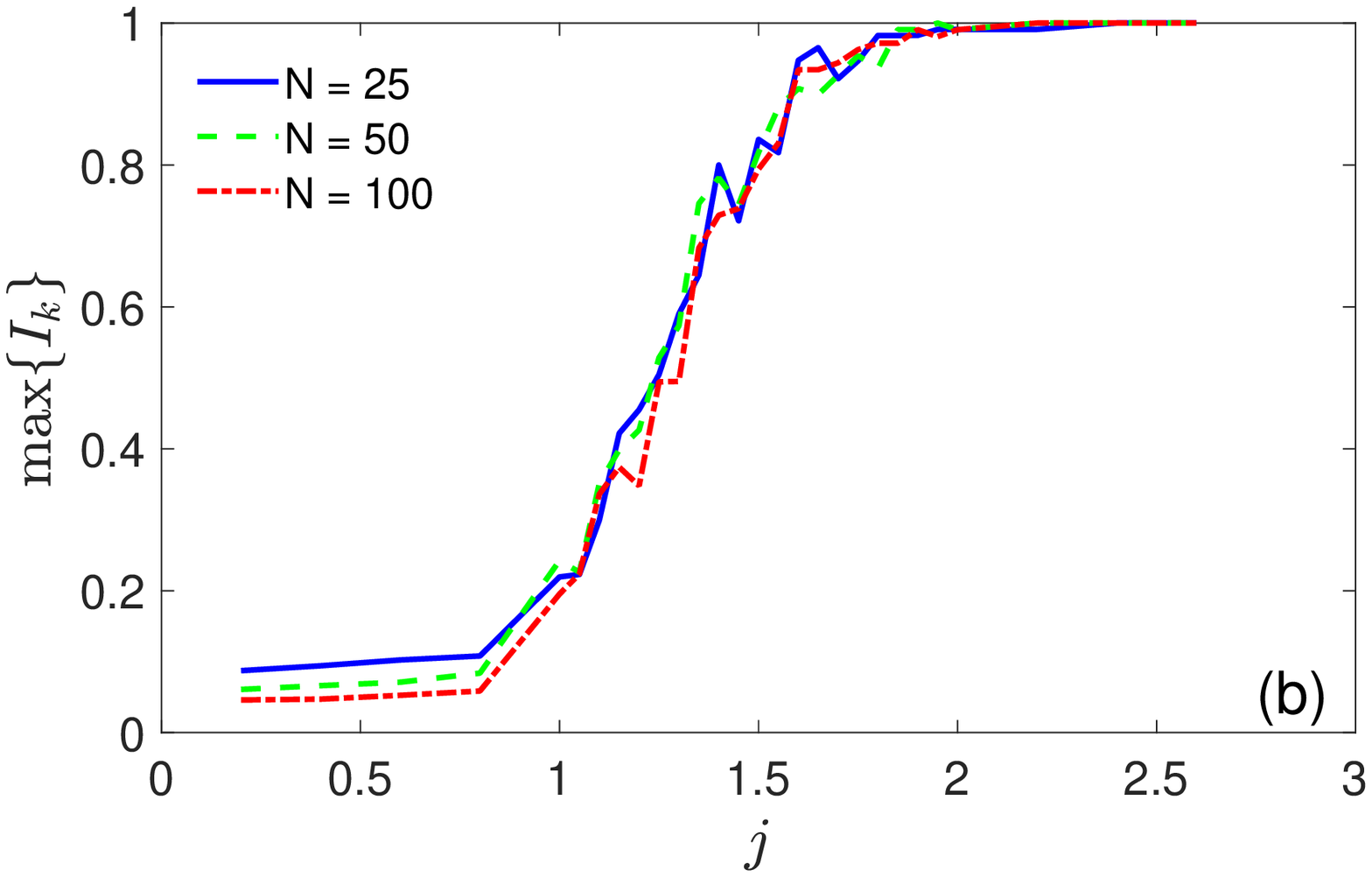}
\end{center}
\caption{
Steady-state value of the maximum $\max_k \{I_k(t)\}$ as a function of
the interaction amplitude $j$.  This quantity was computed by setting
$M = 100$, $T = 50$, three values of $N$ (as indicated in the legend),
and two values of $A$: $A = 1000$ {\bf (a)} and $A = 100$ {\bf (b)}.
}
\label{fig:transition}
\end{figure}

To reduce the dependence on $A$ and $N$, we set $J = j \sqrt{N}/A$,
where $j$ is some intrinsic amplitude of the interactions that is then
rescaled by $A$ and $\sqrt{N}$.  Figure \ref{fig:transition} shows the
steady-state value of the maximum $\max_k \{I_k(t)\}$ as a function of
the interaction amplitude $j$.

\begin{figure}
\begin{center}
\includegraphics[width=65mm]{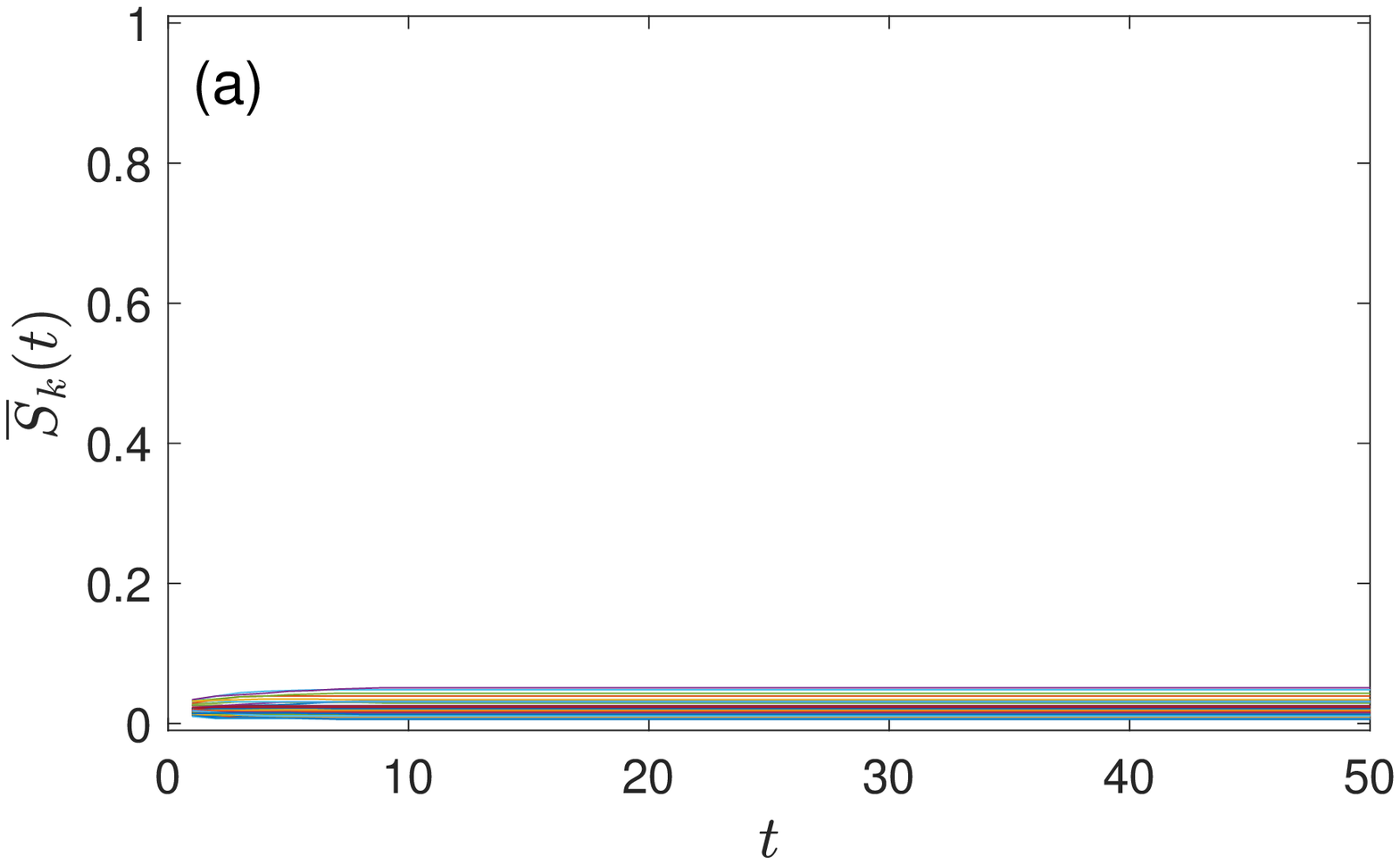}
\includegraphics[width=65mm]{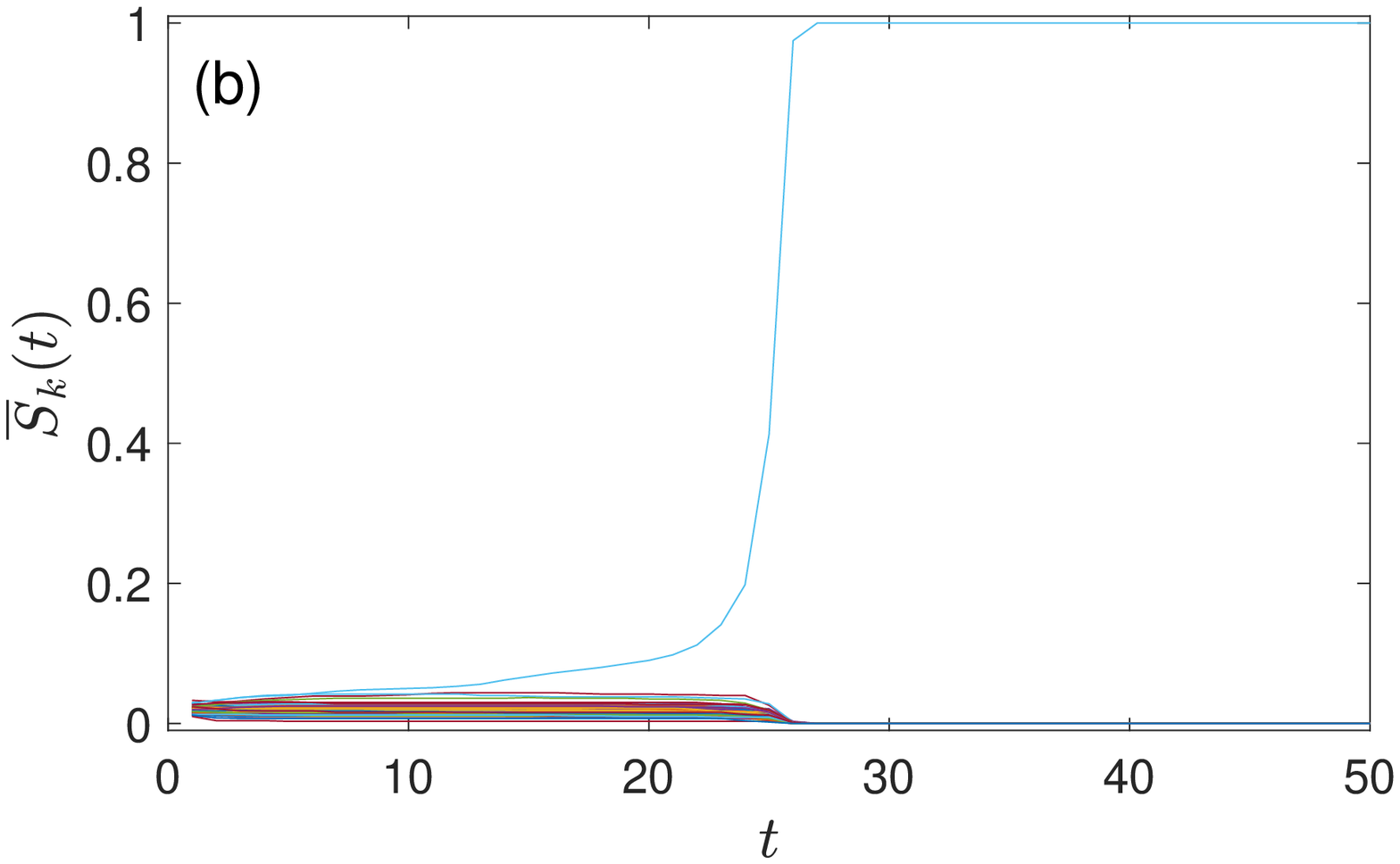}
\end{center}
\caption{
Two random realizations of the dynamics of the fraction of agents,
$\overline{S}_k(t)$, computed for $A = 1000$, $N = 50$, $T = 50$, and
$j = 1.5$, without {\bf (a)} and with {\bf (b)} a dominant strategy.
Fifty curves show time evolution of the fraction $\overline{S}_k(t)$
of each strategy, with $k = 1,2,\ldots, 50$. }
\label{fig:Skt}
\end{figure}

Figure \ref{fig:Skt} shows the dynamics of the fraction of agents,
$\overline{S}_k(t)$, for two simulations.  The choice of an
intermediate level of interactions, $j = 1.5$, leads to two sorts of
outcomes: either there is no dominant strategy (Fig. \ref{fig:Skt}a),
i.e., all strategies remain more or less equally adopted by the
agents; or one dominant strategy emerges (Fig. \ref{fig:Skt}b), while
the remaining strategies are abandonned.  We emphasize that, as all
strategies are equivalent at the beginning, the choice of the
``winner'' strategy is random and realized due to a spontaneous
symmetry breaking among the strategies.  For a smaller level of
interactions (say, $j = 1$ or less), almost all outcomes of
simulations appear without the dominant strategy (not shown).  In
contrast, when interactions are stronger (say, $j = 2$ or higher),
almost all outcomes appear with the dominant strategy.

\begin{figure}
\begin{center}
\includegraphics[width=100mm]{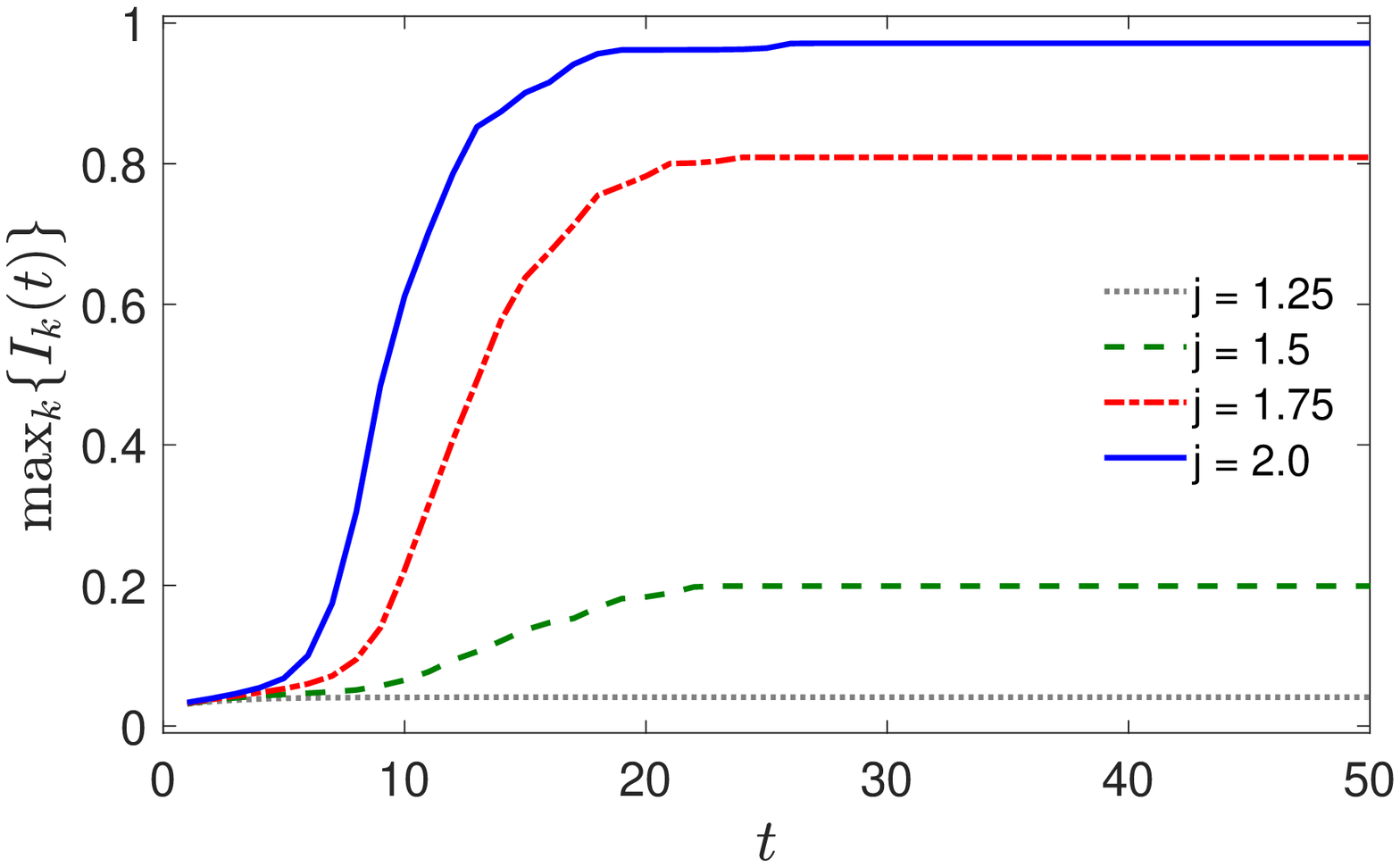}
\end{center}
\caption{
Time evolution of the average maximal fraction of agents interested in
one strategy, $\max_k \{I_k(t)\}$, computed for $A = 1000$, $N = 50$,
$T = 50$, $M = 100$, and four values of the level of interactions $j$
as indicated in the plot.}
\label{fig:maxt}
\end{figure}

Finally, Fig. \ref{fig:maxt} shows the dynamics of the maximal
fraction $\max_k \{I_k(t)\}$.  For weak interactions with $j = 1.25$,
this fraction remains constant, showing that each agent mainly keeps
using its preferred strategy, irrespectively of the others.  At
intermediate interactions ($j = 1.5$), the maximal fraction $\max_k
\{I_k(t)\}$ grows at first time steps and then reaches a 
steady-state value which is larger than in the case $j = 1.25$ but
still relatively small.  This value reflects the fact that some
outcomes do not show a dominant strategy, whereas some other outcomes
do.  As $j$ is further increased, the number of outcomes with the
dominant strategy is getting significantly larger.

In summary, the proposed simplistic model illustrates how interactions
between agents may lead to the emergence of a single dominant strategy
adopted by all agents.  While this model does not aim to mimic or
capture the real mechanisms of decision making in financial trading,
it simply checks that such mechanisms may potentially rationalize the
emergence of a dominant strategy.

\begin{thebibliography}{5}

\bibitem{Clare17}		A. Clare, J. Seaton, P. N. Smith, and S. Thomas, 
				Size matters: Tail risk, momentum, and trend following in international equity portfolios, 
				J. Invest. {\bf 26}, 53-64 (2017).

\bibitem{Covel}			M. W. Covel,
				Trend Following (Updated Edition): Learn to Make Millions in Up or Down Markets,
				Pearson Education, New Jersey, 2009.                 

\bibitem{Clenow}		A. F. Clenow,
				Following the Trend: Diversified Managed Futures Trading,
				Wiley \& Sons, Chichester UK, 2013.

\bibitem{Fung01}		W. Fung and D. A. Hsieh,
				The risk in hedge fund strategies: theory and evidence from trend followers,
				Rev. Financ. Stud. {\bf 14} (2001) 313. 


\bibitem{Clare12}		A. Clare, J. Seaton, P. N. Smith, and S. Thomas, 
				The Trend is Our Friend: Risk Parity, Momentum and Trend Following in Global Asset Allocation, 
				Cass Business School Working Paper (2012); SSRN-id2126478.


\bibitem{Asness13}		C. S. Asness, T. J. Moskowitz, and L. H. Pedersen,
				Value and momentum everywhere,
				J. Finance {\bf 68} (2013) 929. 



\bibitem{Potters05}		M. Potters and J.-P. Bouchaud,
				Trend followers lose more often than they gain,
				Wilmott Magazine (Jan 2006). 

\bibitem{Martin12}		R. Martin and D. Zou,
				Momentum trading: 'skews me, 
				Risk Magazine (2012).




\bibitem{Chan96}		L. K. C. Chan, N. Jegadeesh and J. Lakonishok,
				Momentum Strategies,
				J. Finance {\bf 51} (1996) 1681. 

\bibitem{Jegadeesh01}		N. Jegadeesh and S. Titman,
				Profitability of Momentum Strategies: An Evaluation of Alternative Explanations,
				J. Finance {\bf 56} (2001) 699. 


\bibitem{Moskowitz12}		T. J. Moskowitz, Y. H. Ooi, and L. H. Pedersen,
				Time series momentum,
				J. Finan. Econ. {\bf 104} (2012) 228. 



\bibitem{Hong99}		H. Hong and J. Stein, 
				A unified theory of underreaction, momentum trading and overreaction in asset markets, 
				J. Finan. {\bf 54}, 2143-2184 (1999).

\bibitem{Barberis03}		N. Barberis and R. Thaler, 
				A survey of behavioral finance. 
				In: George M. Constantinides, Milton Harris, and Rene M. Stulz (eds.), 
				{\it The handbook of the economics of finance}, pp. 1053-1128, (2003).

\bibitem{Bhansali15}		V. Bhansali, J. Davis, M. Dorsten, and G. Rennison,
				Carry and Trend in Lots of Places,
				J. Portf. Man. {\bf 41}, 82-90 (2015).

\bibitem{Hurst17}		B. Hurst, Y. H., Ooi, and L. H. Pedersen, 
				A century of evidence on trend-following investing, 
				J. Portf. Man. {\bf 44}, 15-29 (2017).






 

\bibitem{Grebenkov15}		D. S. Grebenkov and J. Serror, 
				Optimal Allocation of Trend Following Strategies, 
				Physica A {\bf 433}, 107-125 (2015).



\bibitem{Benichou16}		R. Benichou, Y. Lemp\'eri\`ere, E. S\'eri\'e, J. Kockelkoren, P. Seager, J.-P. Bouchaud, and M. Potters,
				Agnostic Risk Parity: Taming Known and Unknown-Unknowns,
				J. Invest. Strat. {\bf 6} (3), 1-12 (2017)


\bibitem{Black72}		F. Black, 
				Capital market equilibrium with restricted borrowing, 
				J. Business {\bf 45}, 444-555 (1972).

\bibitem{Asness12}		C. S. Asness, A. Frazzini, and L. H. Pedersen, 
				Leverage aversion and risk parity, 
				Financ. Anal. J. {\bf 68}, 47-59 (2012). 

\bibitem{Chaves11}		D. Chaves, J. Hsu, F. Li, and O. Shakernia, 
				Risk parity portfolios vs. other asset allocation heuristic portfolios,
				J. Invest. 108-118 (Spring 2011).



\bibitem{Grebenkov14}		D. S. Grebenkov and J. Serror,
				Following a Trend with an Exponential Moving Average: Analytical Results for a Gaussian Model,
				Physica A {\bf 394} (2014) 288. 




\bibitem{Thomas12}		S. Thomas, A. Clare, P. N. Smith, and J. Seaton,
				The Trend is Our Friend: Risk Parity, Momentum and Trend Following in Global Asset Allocation.
				Working paper (2012); online: SSRN-id2126478 

\bibitem{Bouchaud}		J.-P. Bouchaud and M. Potters,
				Theory of Financial Risk and Derivative Pricing: From Statistical Physics to Risk Management,
				Cambridge University Press, 2003.

\bibitem{Mantegna}		R. Mantegna and H. E. Stanley, 
				An introduction to Econophysics,
				Cambridge University Press, Cambridge, 1999.

\bibitem{Mantegna95}		R. Mantegna and H. E. Stanley,
				Scaling behaviour in the dynamics of an economic index,
				Nature {\bf 376} (1995) 46. 

\bibitem{Bouchaud01}		J.-P. Bouchaud and M. Potters,
				More stylized facts of financial markets: leverage effect and downside correlations,
				Physica A {\bf 299} (2001) 60.

\bibitem{Sornette03}		D. Sornette,
				Critical Market Crashes,
				Phys. Rep.  {\bf 378} (2003) 1. 

\bibitem{Bouchaud04}		J.-P. Bouchaud, Y. Gefen, M. Potters, and M. Wyart, 
				Fluctuation and response in Financial Markets: the subtle nature of Random price changes,
				Quant. Finance {\bf 4} (2004) 176. 

\bibitem{Bouchaud01b}		J.-P. Bouchaud, A. Matacz, and M. Potters, 
				Leverage effect in financial markets: The retarded volatility model,
				Phys. Rev. Lett. {\bf 87} (2001) 1. 

\bibitem{Valeyre13}		S. Valeyre, D. S. Grebenkov, S. Aboura, and Q. Liu, 
				The Reactive Volatility Model,
                         	Quant. Finance {\bf 13} (2013) 1697. 
\bibitem{Valeyre19}		S. Valeyre, D. S. Grebenkov and S. Aboura, 
The Reactive Beta Model,
Journal of Financial Research {\bf 19} (2019) 71. 

\bibitem{Andersen00}		T. G. Andersen, T. Bollerslev, F. X. Diebold, and P. Labys,
				Exchange Rate Returns Standardized by Realized Volatility are (Nearly) Gaussian,
				Multinat. Finance J. {\bf 4} (2000) 159. 



\bibitem{Hull}			J. C. Hull,
				Options, futures and other derivatives,
				7th Ed., Pearson Prentice Hall, Upper Saddle River, NJ, 2009.




\bibitem{Choueifaty08}		Choueifaty, Y., and Y. Coignard, 
				Toward Maximum Diversification,
				J. Portfolio Management {\bf 35}, 40-51 (2008). 









\bibitem{Black92}		F. Black and R. Litterman,
				Global Portfolio Optimization,
				Finan. Anal. J. {\bf 48}, 28 (1992).

\bibitem{Gadzinski18}		G. Gadzinski, M. Schuller, and A. Vacchino, 
				The Global Capital Stock: Finding a Proxy for the Unobservable Global Market Portfolio,  
				J. Portf. Man. {\bf 44}, 12-23 (2018).

\bibitem{Clarke11}		R. Clarke, H. De Silva, and S. Thorley, 
				Minimum-variance portfolio composition, 
				J. Portf. Man. {\bf 37}, 31-45 (2011).


\bibitem{Nozar88}		H. Nozar and P. Taylor, 
				Stock prices, money supply, and interest rates: the question of causality, 
				Appl. Econ. {\bf 20}, 1603-1611 (1988).

\bibitem{Rahman97}		M. Rahman and M. Mustafa, 
				Dynamic linkages and Granger causality between short-term US corporate bond and stock markets,
				Appl. Econ. Lett. {\bf 4}, 89-91 (1997).


\bibitem{Chan97}		K. C. Chan, S. C. Norrbin, and P. Lai,
				Are stock and bond prices collinear in the long run?
				Int. Rev. Econom. Finance {\bf 6}, 193-201 (1997).

\bibitem{Kasa92}		K. Kasa, 
				Common stochastic trends in international stock markets,
				J. Monet. Econ. {\bf 20}, 196-210 (1992).



\bibitem{Kim02}			C.-J. Kim and J. Piger,
				Common stochastic trends, common cycles, and asymmetry in economic fluctuations,
				J. Monet. Econom. {\bf 49}, 1189-1211 (2002).

\bibitem{Mellander92}		E. Mellander, A. Vredin, and A. Warne,
				Stochastic Trends and Economic Fluctuations in a Small Open Economy,
				J. Appl. Econom. {\bf 7}, 369-394 (1992).

\bibitem{Lee12}			W. Lee,
				Risk On/Risk Off,
				J. Portf. Man. {\bf 38}, 28-39 (2012).




\bibitem{McLean16}		R. D. McLean and J. Pontiff, 
				Does academic research destroy stock return predictability? 
				J. Finan. {\bf 71}, 5-32 (2016).

\bibitem{Valeyre}		S. Valeyre,
				Refined model of covariance/correlation matrix between securities,
				PhD thesis (University Paris 13, France, May 2019).

\bibitem{Hodges17}		P. Hodges, K. Hogan, J. R. Peterson, and A. Ang, 
				Factor Timing with Cross-Sectional and Time-Series Predictors,
				J. Portf. Man. {\bf 44}, 30-43 (2017).



\bibitem{Winters60}		P. R. Winters, 
				Forecasting Sales by Exponentially Weighted Moving Averages,
				Management Science {\bf 6} (1960) 324. 

\bibitem{Brown}			R. G. Brown, 
				Smoothing Forecasting and Prediction of Discrete Time Series,
				Englewood Cliffs, NJ: Prentice-Hall, 1963.







\bibitem{Bun16}			J. Bun,  J.-P. Bouchaud, and M. Potters,
				Cleaning correlation matrices,
				Risk publication (2016).

\bibitem{Bun162}		J. Bun,  J.-P. Bouchaud, and M. Potters,
				Cleaning large Correlation Matrices: tools from Random Matrix Theory,
				Phys. Rep. {\bf 666}, 1-109 (2017).








\bibitem{Michard05}		Q. Michard and J.-P. Bouchaud,
				Theory of collective opinion shifts: from smooth trends to abrupt swings,
				Eur. Phys. J. B {\bf 47}, 151-159 (2005).

\bibitem{Bouchaud13}		J.-P. Bouchaud, 
				Crises and Collective Socio-Economic Phenomena: Simple Models and Challenges,
				J. Stat. Phys. {\bf 151}, 567-606 (2013).

\end{thebibliography}
\end{document}